\documentclass[
superscriptaddress,
     amsmath,amssymb,
     aps,
    pra,
    onecolumn,
    ]{revtex4-2}
\usepackage{tcolorbox}
\usepackage{adjustbox}
\usepackage{xcolor}
\usepackage{graphicx}
\usepackage{tikz} 
\usepackage{dcolumn} 
\usepackage{bm} 
\usepackage{braket} 
\usepackage{amsthm}
\usepackage{bbm}
\usepackage[normalem]{ulem}
\usepackage[ruled]{algorithm2e}
\usepackage{booktabs}
\usepackage{mathtools}
\usepackage{tabularray}

\newtheorem*{theorem-non}{Theorem}
\newtheorem*{lemma-non}{Lemma}
\newtheorem*{corollary-non}{Corollary}
\newtheorem*{proposition-non}{Proposition}
\newcommand{\ntu}{\textsf{NTU}}

\usepackage{appendix} 
\usepackage[colorlinks=true,
            citecolor=NatureTeal,
            linkcolor=NatureRed,
            urlcolor=NatureRed]{hyperref}

\usepackage{fontawesome5}
 
\definecolor{NatureTeal}{HTML}{0F6B68}
\definecolor{NatureBlue}{HTML}{1F4E79}
\definecolor{NatureRed}{HTML}{A23B3B}

\newcommand{\projectpage}[1]{
  \vspace{0.3em}
  \noindent\mbox{
    \textsf{\textcolor{NatureTeal}{\faAtom\ NTU project page:}}~

    \href{#1}{\textcolor{NatureTeal}{\texttt{#1}}}

  }

}

\begin{document}

\title{Efficient foundation decoders for fault-tolerant quantum computing} 

\author{Ge Yan}
\affiliation{College of Computing and Data Science, Nanyang Technological University, Singapore}
\author{Shanchuan Li}
\affiliation{College of Computing and Data Science, Nanyang Technological University, Singapore}
\affiliation{Department of
Electrical Engineering and Computer Science, Tokyo University
of Agriculture and Technology, Koganei, Tokyo, Japan}
\author{Shiyi Xiao}
\affiliation{College of Computing and Data Science, Nanyang Technological University, Singapore}
\affiliation{School of Artificial Intelligence, Shanghai Jiao Tong University, Shanghai, China}
\author{Pengyue Ma}
\affiliation{College of Computing and Data Science, Nanyang Technological University, Singapore}
\author{Hanyan Cao}
\affiliation{Science, Mathematics and Technology Cluster, Singapore University
 of Technology and Design, Singapore}
\author{Feng Pan}
\email{feng\_pan@sutd.edu.sg}
\affiliation{Science, Mathematics and Technology Cluster, Singapore University
 of Technology and Design, Singapore}
\author{Yuxuan Du}
\email{yuxuan.du@ntu.edu.sg}
\affiliation{College of Computing and Data Science, Nanyang Technological University, Singapore}
\affiliation{School of Physical and Mathematical Science, Nanyang Technological University, Singapore}
    
\begin{abstract}
Foundation decoders, a class of high-capacity neural decoders, are leading candidates for fault-tolerant quantum computing, with accurate and efficient decoding at large code distances. However, their construction often faces a steep scaling barrier, as larger code distances rapidly amplify the cost of syndrome generation and neural optimization. To address this bottleneck, here we devise neural transfer unification (\ntu), a unified framework for efficient foundation decoders. A central feature of \ntu\ is its ability to align decoding tasks across code distances via algebraic structures shared by scalable code families, which enables knowledge learned on smaller codes to accelerate large-scale decoder training. We instantiate \ntu\ as \ntu-Transformer, a transformer-based neural decoder tailored for planar surface codes and bivariate bicycle codes. For planar surface codes under circuit-level noise, \ntu-Transformer outperforms correlation-aware matching on the $[\![361,1,19]\!]$ code and further scales to the $[\![625,1,25]\!]$ code, where it exceeds standard matching through transfer adaptation. For the bivariate bicycle code with $[\![72,12,6]\!]$, it surpasses Relay-BP in the low-physical-error regime. These results establish our proposal as a scalable route to amortized cross-distance training of foundation decoders for fault-tolerant quantum processors.

\noindent\projectpage{https://grahamyan.github.io/ntu-decoder}
\end{abstract}

\maketitle

\section{INTRODUCTION}

\begin{figure}
    \centering
    \includegraphics[width=\linewidth]{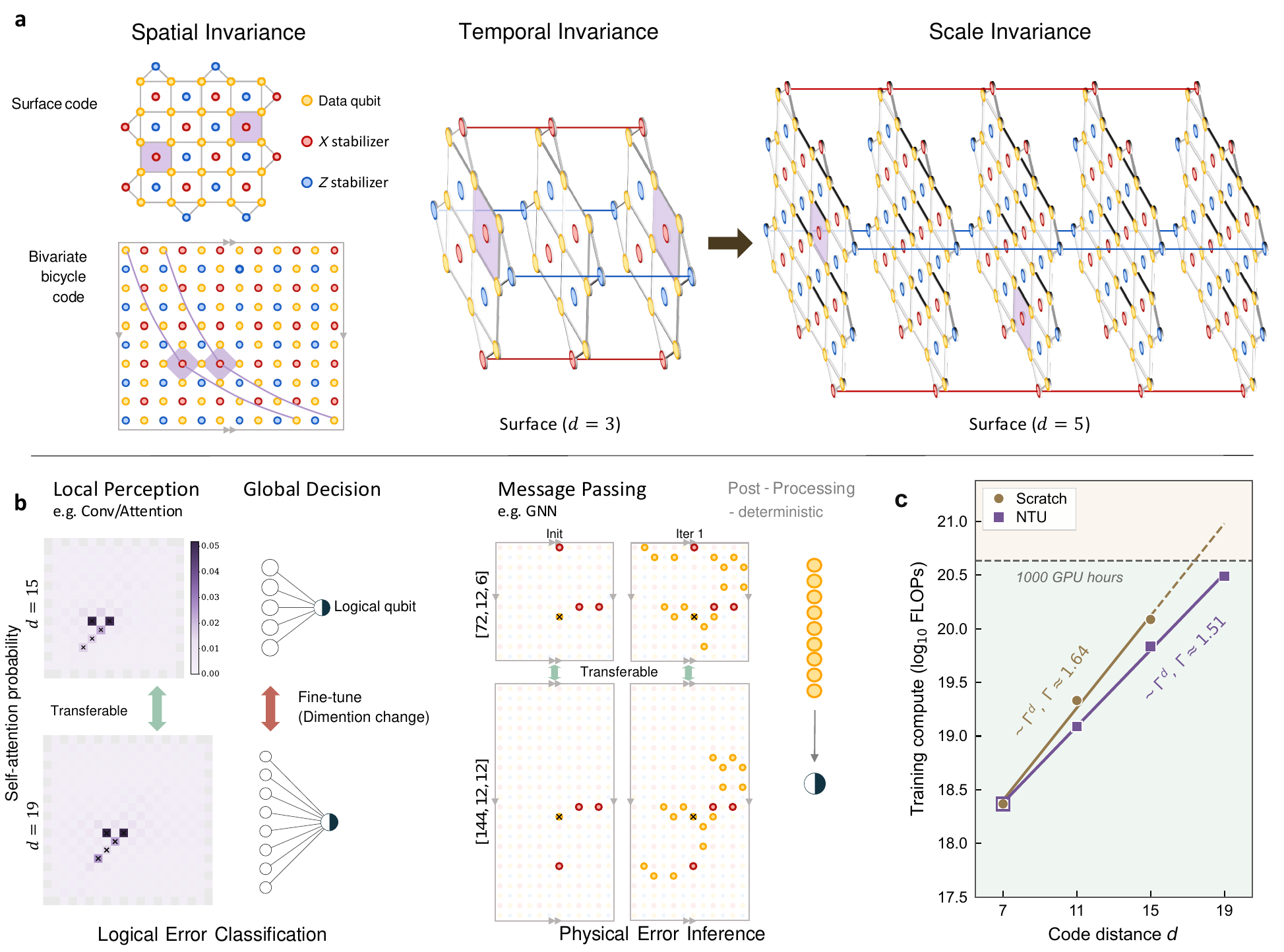}
    \caption{\textbf{QEC code invariance and network transfer.} \textbf{a,} Topological invariances in quantum error correction codes. The fundamental motifs of both planar surface codes and quasi-cyclic bivariate bicycle codes exhibit scale invariance. This scale invariance manifests across spatial translations, temporal syndrome extraction cycles, and large distance expansions. The shaded regions explicitly denote the invariant local topological neighborhoods defined by the codes' generating polynomials. 
    \textbf{b,} Neural decoder network transfer. For logical error classification (left), the architecture is explicitly decoupled into a scale-invariant local perception backbone and a scale-dependent global decision layer. The heatmaps provide a qualitative comparison of self-attention for a matched detector-error pattern on the $d=15$ and $d=19$ lattices. Their similar concentration around matched local relations is consistent with the reuse of relation-aware features after transfer. For physical-error inference models, the transfer mechanism is different: the neural decoder learns local message-passing rules for predicting recovery information, and the final logical parity is then resolved by deterministic post-processing. The stability of these localized rules across code sizes supports the same conclusion that \ntu\ reuses scale-invariant decoding structures rather than relearning distance-specific representations from scratch.
    \textbf{ c,} Computational overhead for scaling to large code distances. The training compute is evaluated as the threshold required to exceed standard PyMatching accuracy. While train-from-scratch methods suffer from severe cold-start plateaus, the \ntu~transfer paradigm fundamentally suppresses the computational scaling exponent  (reducing $\Gamma$ from 1.64 to 1.51), enabling highly efficient network generation at massive code distances.}
    \label{fig:scheme}
\end{figure}

\noindent Fault-tolerant quantum computation is widely envisioned as the route to unlocking the full promise of quantum computing in the presence of pervasive hardware noise~\cite{shor1996fault,steane1999efficient,campbell2017roads}. Central to this is quantum error correction (QEC), which protects logical information by encoding it redundantly across many physical qubits~\cite{gottesman2009introduction,terhal2015quantum,roffe2019quantum}. With physical noise below a threshold, increasing the code distance enables QEC to exponentially suppress logical error rates. This makes the development of QEC architectures that are both powerful and practically implementable a long-standing challenge, owing to the difficulty of balancing robustness, overhead, and compatibility with hardware constraints~\cite{shor1995scheme,steane1996error,gottesman1997stabilizer}. Among existing proposals~\cite{bombin2006topological,kitaev2003fault,dennis2002topological}, surface codes~\cite{dennis2002topological,fowler2012surface,wang2011surface} and bivariate bicycle (BB) codes~\cite{bravyi2024high} stand out as leading candidates, supported by rapid experimental progress across superconducting, trapped-ion, and neutral-atom platforms~\cite{egan2021fault,ryananderson2021realization,google2023suppressing,bluvstein2024logical,bausch2024learning,lacroix2025scaling,google2025quantum}.

Efficient and accurate decoders are indispensable for realizing the full potential of these QEC codes. For this reason, substantial efforts have been made to develop decoding methods, which can be categorized into two broad paradigms: traditional and neural approaches. The former, including maximum-likelihood decoding~\cite{dennis2002topological,bravyi2014efficient,sundaresan2023demonstrating,cao2026maximum}, belief propagation~\cite{mackay2004sparse,poulin2008iterative,yao2024belief}, and matching-based decoders~\cite{edmonds1965paths,wu2023fusion,higgott2025sparse}, have become standard tools in modern QEC experiments because of their high decoding accuracy, but their computational efficiency degrades at large code distances. Neural decoders offer a complementary paradigm by learning decoding rules directly from syndrome data~\cite{torlai2017neural, meinerz2022scalable,ni2020neural,yan2026rethink}, although early models often lagged behind leading traditional decoders in accuracy. \textit{Foundation decoders} have transformed neural decoding into a competitive route for practical QEC. Concrete examples are AlphaQubit series~\cite{bausch2024learning,lacroix2025scaling} and Cascade~\cite{gu2026scalable}, featured by high-capacity architectures and large-scale training. Despite this progress,  current foundation decoders face a severe scaling bottleneck, with training at only modest code distances already requiring thousands of GPU-hours and direct extension to practically relevant large-distance codes becoming untenable. This raises a question: \textit{can foundation decoders be scaled more efficiently?}

In this study, we provide a positive affirmation by introducing neural transfer unification (\ntu), a framework for efficiently constructing foundation decoders via \textit{cross-distance transfer}. Conceptually, \ntu\ exploits scale-invariant decoding structures across code distances, aligning different-sized decoding tasks within a shared representation space. Attributed to this alignment, \ntu\ can transfer the decoding knowledge learned at smaller distances within the same code family to foundation decoders operating at larger code distances. In this regard, \ntu\ departs from the curriculum learning paradigm of current foundation decoders~\cite{bengio2009curriculum}, replacing costly staged training with amortized cross-distance transfer. This paradigm shift turns the foundation decoders from distance-specific training into reusable scaling, substantially reducing computational overhead. 

For practical deployment, \ntu\ can be instantiated with diverse neural architectures. We realize this flexibility as \ntu-Transformer, a transformer-based foundation decoder designed for high-accuracy cross-distance decoding in planar surface codes and BB codes. Its architecture is built around two specific designs: a new scalable transformer embedding model and a geometry-aware rotary positional encoding, which embed variable-size syndrome data into a shared latent representation space. Together, these designs preserve the scale-invariant decoding structures shared across code distances while capturing global syndrome correlations, enabling effective cross-distance transfer. We conduct systematic experiments against strong available baselines to validate the effectiveness and scalability of \ntu-Transformer. For surface codes, \ntu-Transformer outperforms highly optimized correlation-aware matching on the $[\![361,1,19]\!]$ code. Beyond accuracy, \ntu\ substantially reduces the empirical compute scaling of foundation-decoder training, enabling transfer to the larger $[\![625,1,25]\!]$ code, where it achieves lower logical error rates than standard matching without restarting large-scale training. For the [\![72,12,6]\!] BB code, it also outperforms Relay-BP~\cite{maurer2025real} in the low-physical-error regime. Beyond these benchmarks, \ntu-Transformer addresses a broader reproducibility gap in foundation decoding by providing an open and end-to-end pipeline for construction, transfer, and evaluation. By making the full pipeline and trained decoders openly available, our work lowers a key computational barrier to high-accuracy neural decoding and provides a practical starting point for real-time, hardware-integrated decoders in fault-tolerant quantum processors.

\section{RESULTS}
\subsection{Scale invariance in QEC}
A QEC code is conventionally denoted by $[\![n,k,d]\!]$, where $n$, $k$, and $d$ specify the number of physical qubits, logical qubits, and the code distance, respectively. In addition, given a syndrome-extraction circuit, the set of $m_d$ syndrome detectors is denoted by $\mathcal{D}_d$. Among the many quantum code families, Calderbank–Shor–Steane (CSS) codes are particularly prominent and include several leading architectures for fault-tolerant quantum computation~\cite{calderbank1996good,steane1996multiple}. Given their broad relevance, we focus here on structural CSS code families whose larger instances are generated by repeated construction rules. This setting includes planar surface codes and BB codes, and can also extend to other structured quantum low-density parity-check families~\cite{breuckmann2021quantum}, e.g., lifted-product codes~\cite{panteleev2021quantum}, whose generator rules are preserved across distances. 

\begin{figure*}
    \centering
    \includegraphics[width=\linewidth]{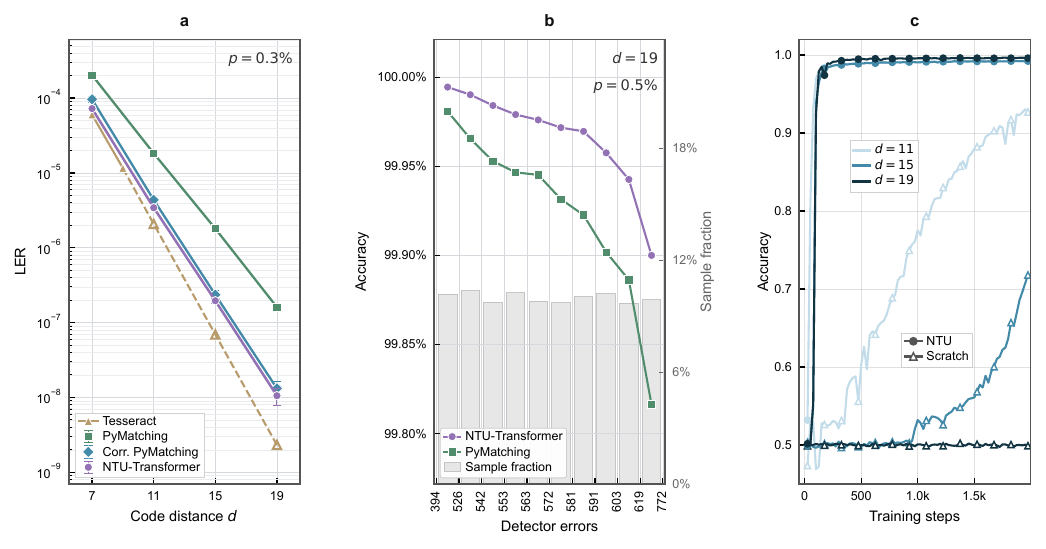}
    \caption{\textbf{Decoding performance and topological transfer dynamics of the neural decoder on the planar surface code. a,} Logical error per round (LER) as a function of code distance $d \in \{7, 11, 15, 19\}$ evaluated at a physical error rate of $p = 0.3\%$. The transferred base model (purple circles) demonstrates highly competitive fault-tolerant scaling, significantly outperforming standard PyMatching (green squares) and closely approaching the performance of Correlated PyMatching (blue diamonds), establishing a robust baseline for large-distance scaling.
    \textbf{b,} Decoding accuracy over $d$-rounds resolved by the number of detector errors within the evaluation window on $d=19$. The underlying histogram denotes the sample fraction distribution across the error weights. While traditional matching-based algorithms experience severe performance degradation in highly degenerate, high-weight error regimes, \ntu-Transformer maintains superior accuracy.
    \textbf{c,} Resolution of the cold-start problem via \ntu. The plot contrasts the training dynamics (accuracy versus fine-tuning steps) of models trained from scratch (dashed lines) against those initialized via \ntu~(solid lines). Without \ntu, models suffer from a prolonged optimization plateau; notably, the $d=19$ model (dashed purple line) remains completely stagnant at 50\% accuracy even after 2,000 steps. In stark contrast, transferred models exhibit instantaneous vertical convergence. This empirical evidence verifies that the local perception backbone perfectly maps to the topological invariance of the larger lattice, leaving only the global decision layer to quickly adapt to the expanded boundaries.}
    \label{fig:surface}
\end{figure*}

We next formulate the scale invariance for these codes, which serves as the organizing principle for cross-distance transfer in \ntu. Intuitively, it refers to preserving the relative neighborhood structure of syndrome detectors as the code distance $d$ changes. To formalize this notion, we describe syndrome detectors in an algebraic language. A detector $v\in \mathcal{D}_d$ is labeled by an algebraic coordinate with $v=x^i y^j t^r$, where $x$ and $y$ encode the repeated spatial or algebraic structure of the code family, $t$ amounts to the repeated syndrome-extraction rounds, and the integers $i$, $j$, and $r$ specify the detector coordinates along these directions. 

For a detector $v$, its neighborhood is determined by the relative shifts from $v$ to the detectors connected to it by the code construction. The allowed relative shifts form a finite set $\mathcal{M}(x,y,t; d)$, i.e., a detector $u\in \mathcal{D}_d$ belongs to the neighborhood of $v$ when $v^{-1}u\in \mathcal{M}(x,y,t; d)$, giving 
\begin{equation}    \label{eq:neighbor}
    \mathcal{N}(v;d)
    =
    v\cdot\mathcal{M}(x,y,t;d).
\end{equation}

For code families considered here, $\mathcal{M}(x,y,t;d)$ is fixed by their repeated construction rules. Refer to Supplementary Information (SI)~\ref{apx:preliminary} for surface and BB cases. As $d$ increases, the detector set $\mathcal{D}_d$ expands, while $\mathcal{N}(v;d)$ for $\forall v$ retains the same relative structure encoded by $\mathcal{M}$. This preserved relative structure enables \ntu\ to transfer decoder representations across code distances.
 
\subsection{Framework of \ntu}
\ntu\ is designed to address the computational barrier in constructing foundation decoders at large code distances. This barrier arises because foundation decoders must optimize many parameters over large syndrome datasets, a cost that grows rapidly with code distance. Existing approaches alleviate this burden through curriculum-like learning~\cite{bengio2009curriculum}, but still build distance-specific representations largely from scratch. As a result, the computational cost grows sharply with $d$ (see SI~\ref{apx:preliminary} for more details). Rather than scaling the curriculum to ever larger distances, \ntu\ formulates foundation-decoder construction as a transfer-learning problem~\cite{pan2010survey}, in which knowledge acquired from smaller distances is structurally aligned and reused at larger distances.

\ntu\ consists of \textit{four steps}: 
spatiotemporal invariance discovery, distance-independent syndrome encoding, neural decoding, 
and transfer adaptation (refer to Methods for details). The first step turns the scale invariance in Eq.~(\ref{eq:neighbor}) into a transferable decoding principle. Rather than treating detectors at different distances as unrelated input coordinates, \ntu\ identifies detector environments generated by the same relative-shift rules and regards them as instances of the same decoding pattern. For example, the surface code in Fig.~\ref{fig:scheme}a repeats the same local check unit as $d$ increases, with red and blue stabilizer checks connected through nearby data qubits. Similarly, a BB code repeats algebraic connectivity patterns generated by cyclic shifts. Syndrome extraction further repeats these check-connectivity rules across measurement rounds. These repeated patterns allow \ntu\ to align equivalent decoding environments and transfer decoding representations across code distances.

The second step converts the invariance identified above into a distance-independent syndrome representation that is shared by neural decoders across distances. Achieving this is challenging because the syndrome $S_d\in\{0,1\}^{m_d}$ is indexed by the detector set $\mathcal{D}_d$, whose size varies with the code distance $d$. Consequently, a standard neural decoder ties its input coordinates and embeddings to a distance-specific detector set, preventing direct reuse across distances. To remove this dependence, \ntu\ represents each matched neighborhood $\mathcal{N}(v;d)$ across code distances by their relative shifts in $\mathcal{M}(x,y,t;d)$. As illustrated in Fig.~\ref{fig:scheme}a, a matched neighborhood for the surface codes is the local syndrome environment around a stabilizer detector, encoded by horizontal, vertical, and temporal offsets rather than by absolute lattice locations. Mathematically, for each detector $v$, \ntu\ constructs an input feature by aggregating the syndrome values in its matched neighborhood, i.e.,
\begin{equation}
    \mathbf{h}(v)
    =
    \sigma \Big(
    \sum_{u\in\mathcal{N}(v;d)}
   \mathbf{w}_{v^{-1}u}\, S_d(u)
    \Big),
    \label{eq:aggregation}
\end{equation}
where $S_d(u)$ is the syndrome value at detector $u$, $\sigma(\cdot)$ is a nonlinear activation, and $\mathbf{w}_{v^{-1}u}\in\mathbb{R}^{q}$ is a learnable embedding vector associated with the relative shift
$v^{-1}u\in\mathcal{M}(x,y,t;d)$. In this way, the input-feature construction yields a distance-independent syndrome encoding that can be reused across code distances.

The third step applies a neural decoder $f(S_d; \bm{\theta})$ to the encoded syndrome $\bm{z}_d=\{\mathbf{h}(v):v\in\mathcal{D}_d\}$ in Eq.~(\ref{eq:aggregation}). 
Here, $\bm{\theta}$ denotes all trainable parameters of the neural decoder, e.g., the embedding parameters $\mathbf{w}_{v^{-1}u}$ and those of the backbone. As a backbone-agnostic framework, \ntu\ can instantiate $f(S_d; \bm{\theta})$ with any neural architecture that processes encoded detector features and their relations, because the distance dependence has already been absorbed into the input encoding. When the goal is to construct a foundation 
decoder, the chosen backbone is taken to be sufficiently high-capacity, with a large number of trainable parameters. Once the decoder architecture is fixed, \ntu\ first trains it on a smaller and tractable code at distance $d$, yielding optimized parameters $\bm{\theta}^*$ for the source decoder $f(S_d; \bm{\theta}^*)$. The fourth step, transfer adaptation, initializes the decoder at a larger distance $d'$ with these parameters, i.e., $\bm{\theta}\leftarrow\bm{\theta}^*$, and fine-tunes it on the encoded syndromes $\bm{z}_{d'}$~\cite{pan2010survey}. Repeating this procedure grows the decoder across distances without restarting large-scale training, completing the \ntu\ framework for amortizing the cost of foundation decoding.

\subsection{Implementation of \ntu-Transformer} 
To build a foundation decoder for both surface codes and BB codes, \ntu\ is instantiated with a newly designed Transformer backbone, dubbed \ntu-Transformer. The proposed decoder has three key components: syndrome-dataset construction, a transfer-optimization procedure, and a QEC-aware Transformer architecture. Below, we briefly describe these ingredients and provide the implementation details in Methods and SI~\ref{apx:transformer}.

The training set at the distance $d$ is denoted by $\mathcal{T}_d=\{(S_d^{(i)},\bm{y}_d^{(i)})\}_{i=1}^{N_d}$, where $S_d^{(i)}$ is the raw syndrome, $\bm{y}_d^{(i)}$ is the supervision label for the chosen decoding task, and the superscript $i$ denotes the $i$-th training example. For surface-code decoding, $\bm{y}_d$ records the induced logical error. For BB-code decoding with physical-error inference, the label records low-level error or recovery variables associated with the decoding graph. The datasets $\mathcal{T}_d$ are generated first at smaller, tractable distances to train the foundation decoder, whereas larger-distance datasets are then generated for transfer adaptation and evaluation.

Once $\{\mathcal{T}_d\}$ are prepared, \ntu-Transformer is trained in two stages. At the first stage, the model is optimized on $\mathcal{T}_d$ at a smaller and tractable distance $d$ using a task-specific supervised objective, i.e.,
\begin{equation}
    \mathcal{L}(\mathcal{T}_d ;\bm{\theta})
    =
    \frac{1}{N_d}
    \sum_{i=1}^{N_d}
    \ell \Big(f \big(S_d^{(i)};\bm{\theta} \big), \bm{y}_d^{(i)}\Big),
    \label{eq:training_loss}
\end{equation}
where the per-sample loss $\ell(\cdot, \cdot)$ is chosen according to the decoding output. Denote the optimized source decoder by $f(S_d; \bm{\theta}^*)$. Then the transfer adaptation initializes the larger-distance decoder with $\bm{\theta}\leftarrow\bm{\theta}^*$ and fine-tunes it on the dataset $\mathcal{T}_{d'}$ with $d'>d$. As specified by the \ntu\ framework, repeating this procedure across distances amortizes the cost of foundation-decoder training without restarting optimization at each larger distance.

The last ingredient is instantiating the decoder $f(S_d;\bm{\theta})$ as \ntu-Transformer, combining standard attention blocks~\cite{vaswani2017attention} with two code-aware components. The first component is the redesigned scalable transformer embedding model~\cite{sadhukhan2026stem}, which implements the distance-independent syndrome encoding $\mathbf{h}$ in Eq.~(\ref{eq:aggregation}) and maps the encoded syndrome $\bm{z}_d$ into Transformer tokens. The second component is a redesigned and QEC-aware rotary position embedding (RoPE)~\cite{su2024roformer}, which supports effective global attention without disrupting detector relations across code distances. The proposed RoPE is essential, as standard positional embeddings are incompatible with the transfer principle of \ntu. Concretely, in curriculum-style Transformer-based decoders~\cite{lacroix2025scaling,bausch2024learning}, the same local detector relation within $\mathcal{N}(v;d)$ can receive distinct positional representations when a surface-code lattice is padded or rescaled to a larger distance $d'$. This mismatch prevents direct cross-distance transfer and contributes to the scaling barrier visualized in Fig.~\ref{fig:scheme}c. To avoid this, \ntu-Transformer applies RoPE to relative algebraic displacements consistent with the shift set $\mathcal{M}(x,y,t;d)$ in Eq.~(\ref{eq:neighbor}). This design preserves detector geometry learned at smaller distances, allowing global attention to transfer to larger codes without relearning distance-specific positional structure.

\smallskip
\noindent\textit{Remark}. As a flexible framework, \ntu\ can be instantiated with different neural decoder backbones. For BB codes, we also instantiate \ntu\ with a neural belief-propagation backbone~\cite{liu2019neural}, dubbed \ntu-Neural-BP. Implementation details are provided in SI~\ref{apx:neural-BP}. 

\begin{figure*}
    \centering
    \includegraphics[width=\linewidth]{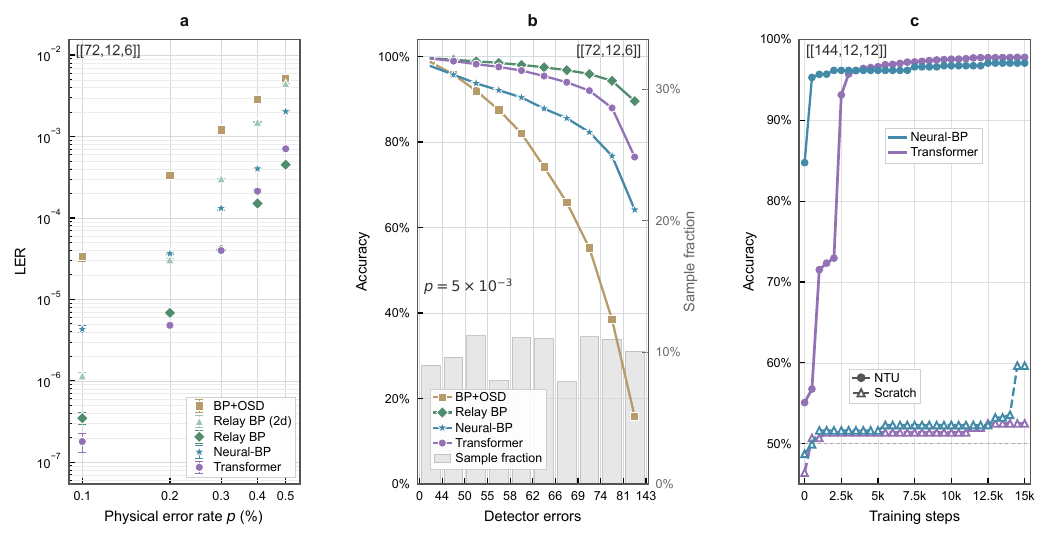}
    \caption{\textbf{Performance evaluation and transfer learning dynamics on non-local BB codes. a,} Logical error rates evaluated on the fundamental $[\![72, 12, 6]\!]$ base code. The \ntu-Transformer strictly outperforms standard BP+OSD and achieves comparable performance against advanced multi-stage RelayBP across all tested physical error rates $p$. \textbf{b,} Decoding accuracy resolved by the number of detector errors at a fixed physical error rate of $p=0.5\%$. The background histogram indicates the sample fraction for each error-weight bin. While traditional heuristics experience sharp performance drops in the high-detector-error regime due to massive syndrome degeneracy, the neural decoder sustains robust accuracy. \textbf{c,} Training convergence on the $[\![144, 12, 12]\!]$ code. Validation accuracy is tracked over the initial training steps to highlight the cold-start phase. Networks initialized from scratch (empty triangles) languish in a prolonged optimization plateau. Conversely, transferring the local perception backbones (solid circles) allows both the \ntu-Transformer and \ntu-Neural-BP to successfully bypass this initial optimization trap, yielding drastically accelerated convergence.}
    \label{fig:BB}
\end{figure*}

\section{Experimental results}
We next conduct systematic experiments on surface codes and BB codes to evaluate the decoding performance and cross-distance transferability of \ntu. Additional implementation details and extended simulation results are provided in {SI~\ref{apx:exp_details}\&\ref{apx:results}}.

\subsection{Surface codes}

The first task applies \ntu-Transformer to planar rotated surface codes, $[\![d^2,1,d]\!]$, using a standard $Z$-basis memory experiment~\cite{google2023suppressing,google2025quantum}. To understand how \ntu-Transformer scales as a foundation decoder, we increase the code distance $d$ from $7$ to $19$ ($97$--$721$ physical qubits) and compare it with three leading decoders: minimum-weight perfect matching (MWPM) using PyMatching, its correlated-noise variant (MWPM-Corr), and Tesseract~\cite{higgott2022pymatching,higgott2023improved,beni2025tesseractdecoder}. The logical error per round (LER)~\cite{google2023suppressing}, i.e., the fraction of experiments in which the decoder would fail per syndrome-extraction round, measures the performance. Unless otherwise specified, all experiments use $R=d$ syndrome-extraction rounds, with samples generated by Stim~\cite{gidney2021stim} under a circuit-level depolarizing noise model with uniform physical error rate $p$.

Following the transfer pipeline introduced above, \ntu-Transformer contains approximately 32 million trainable parameters. For $d=7,11,15,19$, it is trained with $4.1\times10^7$, $5.7\times10^{7}$, $1.3\times10^{8}$, and $2.9\times10^{8}$ samples under $10^{4}$, $1.4\times10^{4}$, $3.2\times10^{4}$, and $7.1\times10^{4}$ iterations, respectively.

The scaling of the LER of \ntu-Transformer with code distance $d$ at $p=0.3\%$ is shown in Fig.~\ref{fig:surface}a. Evaluation is performed on on-the-fly test examples, where each syndrome sample is generated during evaluation rather than drawn from a fixed pre-generated test set. In this setting, \ntu-Transformer outperforms both MWPM-Corr and MWPM across the tested distances. For example, at $d=19$, it reaches an LER of $1.1\times10^{-8}$, compared with $1.3\times10^{-8}$ for MWPM-Corr and $1.6\times10^{-7}$ for MWPM. Although Tesseract can reach higher accuracy, it relies on hypergraph-based shortest-path heuristics whose computational cost becomes increasingly demanding at larger code distances. We therefore use its extrapolated results beyond 
$d=9$ as a reference rather than as a scalable decoding baseline. These results establish \ntu-Transformer as a scalable foundation decoder for high-accuracy surface-code decoding at large distances.

To further characterize the decoding performance in the high-detector-weight regime, we perform a refined analysis at $d=19$. To this end, we set the physical error rate to $p=0.5\%$, which increases the occurrence of higher-weight detector-error events and more complex syndrome patterns. Figure~\ref{fig:surface}b reports the decoding accuracy as a function of detector-error weight, using  $7.5\times10^{6}$ on-the-fly syndrome samples with comparable fractions across the tested weight bins. Across all weights, \ntu-Transformer achieves higher decoding accuracy over $d$ rounds than MWPM, and the advantage becomes more pronounced in the high-weight regime. For example, at weight $772$, \ntu-Transformer reaches an accuracy of approximately $99.90\%$, compared with $99.82\%$ for MWPM. These results indicate that \ntu-Transformer captures non-local syndrome correlations that are essential for high-accuracy decoding at large code distances, attributed to the tailored global attention mechanism.

We next examine whether \ntu\ reduces the computational barrier of train-from-scratch foundation decoding. To this end, we train the same \ntu-Transformer architecture either with sequential transfer adaptation or from scratch, using matched hyperparameters and training budgets. Figure~\ref{fig:surface}c shows that \ntu-Transformer reaches near-saturated accuracy (i.e., $>99\%$) within $750$ steps for $d=11, 15, 19$. In contrast, scratch training enters a cold-start plateau that worsens with distance: even after another 1250 steps, it reaches only $92.7\%$, $71.8\%$, and $50.0\%$ accuracy at $d=11$, $15$, $19$, respectively. This gap identifies cold-start optimization as a key barrier in large-distance neural decoding and shows that \ntu\ alleviates it through cross-distance transfer.

\subsection{BB codes}

The second task applies \ntu\ to BB codes, extending the evaluation beyond planar geometries. To provide a comprehensive comparison, we benchmark against standard BP+OSD~\cite{roffe2020decoding} and representative Relay-family decoders~\cite{muller2025improved}, including Relay-BP and Relay-2D. As in the surface-code experiments, all experiments use $R=d$ syndrome-extraction rounds, with samples generated under a circuit-level depolarizing noise model with uniform physical error rate $p$. Decoder performance is again reported as LER. For BB codes, a logical failure is counted when the residual error after applying the predicted recovery induces a non-trivial logical operator. To systematically assess the effect of the neural backbone, we evaluate both \ntu-Transformer and \ntu-Neural-BP on the same BB-code decoding tasks.

We begin by benchmarking \ntu\ on the $[\![72,12,6]\!]$ BB code across all noise levels. Each \ntu\ decoder uses a fixed architecture and parameter count across all test configurations, approximately $3.6 \times 10^7$ for \ntu-Transformer and $3.0 \times 10^5$ for \ntu-Neural-BP. For each physical error rate $p\in\{0.1,0.2,0.3,0.4,0.5\}\%$, the models are trained separately for $1.5\times10^4$ update steps, corresponding to $6.1\times10^7$ and $3.1\times10^7$
on-the-fly syndrome samples for \ntu-Transformer and \ntu-Neural-BP, respectively. The results are shown in Fig.~\ref{fig:BB}a. The two \ntu-based decoders exhibit distinct behavior: \ntu-Neural-BP improves over standard BP+OSD, but still shows a noticeable gap to the stronger Relay-family baselines. When $p=0.1\%$, the LERs are $4.3\times10^{-6}$, $1.2\times10^{-6}$, and $3.5\times10^{-7}$ for \ntu-Neural-BP, Relay-2d and Relay-BP, respectively. In contrast, \ntu-Transformer outperforms all reference decoders under small tested physical error rates, reaching an LER of $1.8\times10^{-7}$ with a standard deviation of {$8.5\times10^{-8}$} at $p=0.001$.

We then probe the decoding mechanism of \ntu-Transformer and \ntu-Neural-BP on the $[\![72,12,6]\!]$ BB code. As with the surface-code analysis, Fig.~\ref{fig:BB}b reports decoding accuracy as a function of detector-error weight at $p=0.5\%$, evaluated over $10^6$ on-the-fly syndrome samples. Standard BP-based decoders such as BP+OSD degrade more rapidly in the high-weight regime, where syndrome patterns become more degenerate and correlated. By contrast, \ntu-Transformer maintains good performance across the tested weight bins, e.g., it reaches $97.8\%$ accuracy at detector-error $[81,143)$, compared with $85.7\%$ for BP+OSD.
This indicates that \ntu-Transformer captures non-local syndrome correlations induced by the quasi-cyclic structure of BB codes, extending its high-weight robustness beyond planar geometries.

We lastly examine whether the scale invariance exploited by \ntu\ reduces the training barrier when the BB code is scaled. Figure~\ref{fig:BB}c illustrates the training dynamics when transferring from the $[\![72,12,6]\!]$ BB code to the larger $[\![144,12,12]\!]$ BB code. With random initialization, both neural backbones enter a cold-start plateau, reaching only $52.5\%$ accuracy for \ntu-Transformer and $59.7\%$ accuracy for \ntu-Neural-BP after 1500 training steps. By contrast, \ntu\ initializes the larger-code decoder from the optimized $[\![72,12,6]\!]$ decoder and rapidly reaches high decoding accuracy: \ntu-Transformer reaches $93.1\%$ accuracy within 2500 steps, and \ntu-Neural-BP reaches $95.3\%$ accuracy within 500 steps. The effect appears for both neural backbones, indicating that \ntu\ reuses algebraically matched decoding representations rather than relying on a backbone-specific shortcut. These results show that \ntu\ mitigates the large-distance training barrier across both planar and quasi-cyclic quantum memory architectures.

\section{DISCUSSION}
We have presented \ntu, a novel transfer-learning framework for constructing foundation decoders that scale across code distances. By exploiting scale invariance in scalable code families, \ntu\ converts distance-dependent syndrome data into reusable decoding representations, replacing repeated large-distance optimization with cross-distance transfer. This moves neural decoding beyond prior curriculum-like learning and directly addresses the scaling barrier of foundation decoding. Instantiated as \ntu-Transformer, the framework achieves high-accuracy decoding for planar surface codes up to $[\![361,1,19]\!]$ and extends to BB codes, outperforming leading baseline decoders while substantially reducing computational overhead. These results suggest a shift from distance-specific neural decoders towards reusable foundation decoders for fault-tolerant quantum computation.

Our study points to broader research directions in which AI can advance scalable foundation decoding and create new opportunities for quantum error correction \cite{wang2024artificial,alexeev2025artificial,du2025artificial}. Several important avenues remain to be explored. The first is to identify scale invariance in broader families of quantum codes and design corresponding \ntu\ backbones. Beyond surface and BB codes, many promising code families, such as hypergraph-product codes~\cite{panteleev2021quantum}, lifted-product codes~\cite{breuckmann2021quantum}, and Floquet codes~\cite{hastings2021dynamically,davydova2023floquet}, possess algebraic, geometric, or temporal structures that may be reused across code sizes. Extending \ntu\ to these structures would broaden foundation decoding from the code families studied here to a wider class of scalable QEC architectures. 

Moreover, it is crucial to explore how foundation decoders can be used to achieve real-time decoding~\cite{skoric2023parallel,campbell2024series,barber2025real}. A natural route is to compress high-capacity foundation decoders into lightweight models through hardware-aware techniques such as knowledge distillation, quantization, and network pruning~\cite{cheng2018model}. However, these model-compression techniques rely on powerful foundation decoders that provide reliable supervision signals and reveal the relevant syndrome structure~\cite{yan2026rethink,gu2026scalable}. To the best of our knowledge, \ntu\ is the first open-source foundation decoder for QEC, providing a practical starting point for systematic real-time decoder optimization.

Another important direction is to build foundation decoders for logical circuits beyond quantum memory experiments~\cite{zhou2025opportunities}. Recent progress has begun to move neural decoding from memory benchmarks towards circuit-level logical operations~\cite{zhou2025learning}. However, prior studies remain limited to relatively small code distances, fixed logical gate sets, and specific circuit families, and their accuracy--runtime trade-off still depends on task-specific training and architecture design. This leaves open how to construct foundation decoders that generalize across logical operations, code distances, and syndrome-extraction schedules. \ntu\ with a circuit-aware transfer could address this limitation and provide a unified neural backbone for fault-tolerant quantum processors.

\clearpage

\onecolumngrid
\appendix

\setcounter{figure}{0}
\setcounter{table}{0}
\renewcommand{\thefigure}{\arabic{figure}}
\renewcommand{\thetable}{\arabic{table}}
\makeatletter
\renewcommand{\fnum@figure}{SI~FIG.~\thefigure}
\renewcommand{\fnum@table}{SI~Table~\thetable}
\makeatother
\renewcommand{\appendixname}{SI} 
\makeatother

\tableofcontents

\bigskip
The organization of the Supplementary Information (SI) is as follows. In SI~\ref{apx:preliminary}, we briefly recap the necessary background on surface and BB codes, their
classical and neural decoders, and transfer learning. After that, we elaborate on the 
\ntu\ framework in SI~\ref{apx:NTU}. The implementation details of
its two instantiations, \ntu-Transformer and \ntu-Neural-BP, are provided in SI~\ref{apx:transformer} and~\ref{apx:neural-BP}, respectively. 
SI~\ref{apx:exp_details} specifies the evaluation metrics, training protocols, and
baselines. 
SI~\ref{apx:results} presents more simulation results.

\section{Preliminary}\label{apx:preliminary}

This section collects the background used throughout the SI. Specifically, SI~\ref{subsec:pre_surface} and SI~\ref{subsec:pre_bb} briefly introduce the rotated surface code and bivariate bicycle codes, together with their conventional decoders. SI~\ref{subsec:pre_neural} then reviews neural decoders for these two code families, and SI~\ref{subsec:pre_transfer} summarizes the transfer-learning ingredients that \ntu\ builds on.

\subsection{Surface code}\label{subsec:pre_surface}
\smallskip
\noindent\textbf{Code construction.}
The rotated surface code encodes a single logical qubit, $k=1$, in $n=d^2$ data qubits on a two-dimensional lattice, where $d$ is the code distance. As shown in Fig.~\ref{fig:scheme}a, weight-four $X$- and $Z$-type stabilizers tile alternating plaquettes in a checkerboard pattern, so each bulk check couples only to nearby data qubits. This planar locality makes surface codes well suited to superconducting hardware with limited qubit connectivity. However, it also limits their encoding efficiency: increasing the code distance requires the number of data qubits to grow quadratically.

For a distance-$d$ memory experiment, the syndrome-extraction circuit is repeated for $R$ rounds. Each detector $v\in\mathcal{D}_d$ is a binary parity of selected measurement outcomes, typically comparing stabilizer measurements across rounds, and is assigned a spacetime coordinate $v=x^i y^j t^r$, where $x$ and $y$ index the lattice and $t$ indexes the measurement round. Thus, $\mathcal{D}_d$ spans a three-dimensional syndrome volume with $m_d=|\mathcal{D}_d|=\mathcal{O}(d^2R)$ detectors, and decoding requires reasoning over this full spacetime volume rather than over a single round.

\smallskip
\noindent\textbf{Conventional decoders.} Conventional matching-based decoders operate on a detector graph derived from $\mathcal{D}_d$ and the detector error model (DEM)~\cite{gidney2021stim}. In this graph, nodes correspond to detectors in $\mathcal{D}_d$, and edges correspond to graphlike fault mechanisms. A detector fires in a given shot when its associated measurement parity is nonzero. The DEM further specifies the edge weights and logical-observable effects used by the decoder to infer a correction.

Matching-based decoding then proceeds as follows. From the multi-round syndrome $S_d\in\mathbb{F}_2^{m_d}$, with one bit per detector and $m_d=|\mathcal{D}_d|$, the decoder returns an estimated fault-edge vector $\hat{\bm{e}}\in\mathbb{F}_2^{M}$, where $M$ is the number of graphlike fault edges considered by the decoder. The associated logical prediction is obtained from the logical-observable map $L\in\mathbb{F}_2^{1\times M}$, a fixed binary row vector specified by the same DEM, i.e.,
\[
\hat{\bm{y}}_d = L\hat{\bm{e}}\in\{0,1\}.
\]
The true label $\bm{y}_d\in\{0,1\}$ is the logical observable value produced by the simulator for the same shot. A shot is counted as a logical failure when $\hat{\bm{y}}_d\neq\bm{y}_d$, and averaging this event over shots, followed by the per-round rescaling in SI~\ref{subsec:eval_metrics}, gives the logical error per round (LER). The baselines below differ only in how they choose $\hat{\bm{e}}$, and all share the readout $\hat{\bm{y}}_d=L\hat{\bm{e}}$ (Fig.~\ref{fig:surface}a).

\smallskip
\noindent\emph{MWPM.} Minimum-weight perfect matching (MWPM) pairs detection events through paths on the spacetime detector graph, with edge weights determined by the log-likelihoods of the corresponding fault mechanisms. The decoder returns the minimum-weight set of paths, including possible boundary connections, that explains the observed syndrome~\cite{edmonds1965paths}. This path set defines $\hat{\bm{e}}$, from which the predicted logical flip is read out as $\hat{\bm{y}}_d=L\hat{\bm{e}}$. Here we implement MWPM by using the sparse PyMatching backend~\cite{higgott2022pymatching,higgott2025sparse}.

\smallskip
\noindent\emph{Correlated MWPM.} A single circuit-level fault, such as a $Y$ error, can flip both $X$- and $Z$-type detectors, whereas plain MWPM treats the corresponding matching problems independently. Correlated MWPM (MWPM-Corr) incorporates these cross-type correlations by reweighting edges before a second matching pass~\cite{higgott2023improved}. The resulting matched paths define $\hat{\bm{e}}$, and the logical prediction is again obtained from $L\hat{\bm{e}}$. This improves accuracy while remaining within the matching paradigm.  Here we implement MWPM-Corr by using the sparse Correlated PyMatching backend~\cite{higgott2022pymatching,higgott2025sparse}.

\smallskip
\noindent\emph{Union-find.} Union-find replaces global matching by growing and merging clusters of detection events on the same spacetime detector graph until each cluster can be locally corrected or connected to a boundary~\cite{delfosse2021almost,huang2020fault}. The union of these local corrections defines $\hat{\bm{e}}$, and the logical prediction is computed as before. Union-find has near-linear runtime and low latency, trading accuracy for speed relative to optimized matching.

\subsection{Bivariate bicycle codes}\label{subsec:pre_bb}
\smallskip
\noindent\textbf{Code construction.}
Bivariate bicycle (BB) codes are quantum low-density parity-check (qLDPC) codes built over the group ring $\mathcal{R}=\mathbb{F}_2[x,y]/\langle x^\ell-1,y^m-1\rangle$, where $\ell$ and $m$ specify the two cyclic dimensions of the code, and $x$ and $y$ act as translations along them~\cite{bravyi2024high}. Two sparse polynomials $A,B\in\mathcal{R}$ define the CSS check matrices
\begin{equation}
H_X=[A\ B],\qquad H_Z=[B^{T}\ A^{T}],
\label{eq:bb_matrix}
\end{equation}
where the transpose denotes the group-ring involution. Because the polynomial shifts commute over $\mathcal{R}$, this construction satisfies the CSS condition $H_XH_Z^{T}=0$ over $\mathbb{F}_2$. The rows of $H_X$ and $H_Z$ define the $X$- and $Z$-type stabilizer checks, respectively. For the BB codes considered here, the generators $A$ and $B$ each contain three monomials, so each row of $H_X$ and $H_Z$ has weight six. Thus, unlike the weight-four local checks of the surface code in SI~\ref{subsec:pre_surface}, BB checks are sparse but connect qubits through non-local cyclic shifts.

This algebraic construction gives BB codes a different scaling structure from planar surface codes. Whereas a rotated surface code encodes one logical qubit using $n=d^2$ data qubits, the BB instances studied here encode multiple logical qubits, including the $[\![72,12,6]\!]$ and $[\![144,12,12]\!]$ codes, using $n=2\ell m$ data qubits. Increasing the cyclic dimensions $\ell$ and $m$ enlarges the code while keeping the polynomial generators fixed. As shown in Fig.~\ref{fig:scheme}a, the relative algebraic relations that define the detector neighborhoods repeat across sizes, providing the BB analogue of the scale-invariant decoding structure used by \ntu.

\smallskip
\noindent\textbf{Conventional decoders.} BB-code decoding is naturally formulated on a sparse spacetime Tanner graph derived from the circuit-level DEM. One node set corresponds to detectors $\mathcal{D}_d$ across all syndrome-extraction rounds, and the other to $M$ circuit-level fault mechanisms; an edge indicates that a fault mechanism flips the corresponding detector. We represent this graph by the detector–fault incidence matrix $H_{\mathrm{DEM}}\in\mathbb{F}_2^{m_d\times M}$. Multiplication by $H_{\mathrm{DEM}}$ maps a fault vector to the detector syndrome it would produce, so decoding seeks an estimate $\hat{\bm e}\in\mathbb{F}2^M$ satisfying $H_{\mathrm{DEM}}\hat{\bm e}=S_d$ for the observed multi-round syndrome $S_d\in\mathbb{F}_2^{m_d}$. The graph inherits the sparse algebraic structure of $H_X$ and $H_Z$ in Eq.~(\ref{eq:bb_matrix}), while also incorporating timelike detector relations and measurement faults across syndrome-extraction rounds.

The logical prediction is obtained from the DEM logical-observable map, $\hat{\bm{y}}_d=L\hat{\bm{e}}$, where $L\in\mathbb{F}_2^{k\times M}$ has one row per logical observable. For the $[\![72,12,6]\!]$ BB code, $k=12$, so $\hat{\bm{y}}_d\in\mathbb{F}_2^{12}$ predicts all logical observables simultaneously. Because multiple logical qubits are encoded, we report the per-round, per-logical-qubit logical error rate $\epsilon_{\mathrm{LQ}}$ defined in SI~\ref{subsec:eval_metrics}. The BB baselines in Fig.~\ref{fig:BB}a differ in how they estimate $\hat{\bm{e}}$, while sharing this logical readout.

\smallskip
\noindent\emph{Belief propagation.} Belief propagation (BP) iterates check-to-variable and variable-to-check messages on the DEM Tanner graph to estimate the posterior reliability of each fault mechanism~\cite{mackay2004sparse,poulin2008iterative,roffe2020decoding}. A hard decision on these reliabilities gives an estimated fault vector $\hat{\bm{e}}$, from which the logical prediction is read out as $\hat{\bm{y}}_d=L\hat{\bm{e}}$. The cost scales with the number of message-passing iterations and the number of graph edges. In qLDPC codes, short cycles in the Tanner graph and stabilizer degeneracy can prevent standard BP from producing a syndrome-consistent estimate, motivating the post-processing methods below.

\smallskip
\noindent\emph{Ordered statistics decoding.} BP with ordered statistics decoding (BP+OSD) uses the soft reliabilities produced by BP to select a reliable information set and then solves the syndrome equation on a full-rank subset of columns of $H_{\mathrm{DEM}}$, optionally searching low-order flips among the least reliable fault mechanisms~\cite{roffe2020decoding,panteleev2021degenerate}. This post-processing can restore a syndrome-consistent estimate $\hat{\bm{e}}$ when raw BP fails, after which the logical prediction is again obtained as $\hat{\bm{y}}_d=L\hat{\bm{e}}$. Its cost increases with the OSD order.

\smallskip
\noindent\emph{Local subgraph decoding.} Local subgraph decoding (LSD) localizes the ordered-statistics search to connected subgraphs associated with the residual syndrome, rather than applying a global OSD step to the full Tanner graph~\cite{panteleev2021degenerate}. When the residual syndrome decomposes into localized components, this produces a syndrome-consistent estimate $\hat{\bm{e}}$ at lower cost than global OSD, with the logical readout computed as before.

\smallskip
\noindent\emph{RelayBP.} RelayBP retains the Tanner-graph message-passing structure but augments standard BP with a multi-stage relay procedure. A global BP stage is followed by multiple relay branches with perturbed memory strengths, which helps escape degenerate fixed points that can trap standard BP on qLDPC Tanner graphs~\cite{muller2025improved}. The first syndrome-consistent candidate produced by these relay branches defines $\hat{\bm{e}}$, and the logical prediction is obtained from $L\hat{\bm{e}}$. With sufficient iterations and relay branches, RelayBP provides a strong conventional baseline for BB-code decoding.

\subsection{Neural decoders and foundation decoders}\label{subsec:pre_neural}
\smallskip
\noindent\textbf{Neural decoding paradigms.}
Neural decoders learn a syndrome-to-error or syndrome-to-logical map from data~\cite{torlai2017neural,krastanov2017deep,varsamopoulos2017decoding}. Following the main text, we write such a decoder as a trainable map $f(S_d;\bm{\theta})$ with parameters $\bm{\theta}$, acting on the binary detector history $S_d\in\{0,1\}^{m_d}$ defined in Eq.~(\ref{eq:aggregation}). In our setting, the learning target falls into two paradigms: logical error classification and physical error inference. The former trains the network to predict the logical flip directly from the syndrome, whereas the latter trains it to infer physical-error or recovery information before applying a deterministic logical readout. These two learning paradigms correspond to the surface-code and BB-code experiments described below, respectively.

\smallskip
\noindent\emph{Logical error classification.}
In this paradigm, the neural decoder $f(S_d;\bm{\theta})$ predicts the logical flip directly,
\begin{equation}
\hat{\bm{y}}_d=f(S_d;\bm{\theta})\in\mathbb{F}_2^{k},
\label{eq:lec}
\end{equation}
where $\bm{y}_d$ is the logical label associated with the observed detector history. This target does not require the decoder to identify a particular fault configuration $\hat{\bm{e}}$ as introduced in SI~\ref{subsec:pre_surface} and \ref{subsec:pre_bb}, since many physically distinct recovery chains can have the same logical effect. Instead, the neural decoder must learn which syndrome features are relevant to the final logical parity, effectively collapsing degenerate recoveries into the same classification target. This makes logical error classification a natural formulation for surface-code memory experiments, where the main task is to decide whether a logical observable has flipped.

However, this formulation also imposes an architectural constraint. As the code distance increases, the detector history $S_d$ grows with $|\mathcal{D}_d|$, whereas the output dimension remains fixed by the number of logical observables. A decoder built from flattened distance-specific inputs therefore couples local syndrome recognition to a particular global detector layout, making the model size and input structure explicitly distance dependent. A scalable classifier should instead separate local syndrome feature extraction from the final logical decision stage, so that changes in the syndrome volume do not require redefining the entire model.

\smallskip
\noindent\emph{Physical error inference.}
In this paradigm, the neural decoder is trained to infer a recovery or fault estimate before applying a fixed logical readout. Formally,
\begin{equation}
\hat{\bm{e}}=f(S_d;\bm{\theta})\in\mathbb{F}_2^{M},
\quad \text{with} \quad
\hat{\bm{y}}_d=L\hat{\bm{e}},
\label{eq:pei}
\end{equation}
where the $M$ output coordinates correspond to the fault mechanisms of the decoding graph, and $L$ is the DEM logical-observable map defined in SI~\ref{subsec:pre_surface} and SI~\ref{subsec:pre_bb}. This formulation is naturally aligned with Tanner-graph message passing, because the network can associate learnable updates with detector–fault relations and then use the fixed map $L$ to obtain the logical prediction. It therefore provides a natural setting for neural belief-propagation decoders~\cite{liu2019neural,maan2025machine,bhave2025hypernq}.

However, physical error inference also introduces a degeneracy issue. For a given detector syndrome $S_d$, there are generally many fault estimates $\hat{\bm{e}}$ that reproduce the syndrome and have the same logical effect. Thus, unlike logical error classification, the supervised target for the neural decoder $f$ is not unique at the level of physical or recovery configurations. A decoder based on this paradigm must therefore learn a syndrome-consistent representative whose logical class is correct, rather than a unique underlying error.

\smallskip
\noindent\textbf{Neural architectures.}
The choice of neural architecture (or neural backbone) is orthogonal to the choice of learning paradigm. The two learning paradigms above specify what the neural network $f$ is trained to predict, whereas the architecture specifies how the detector history $S_d$ is represented and processed in $f$. 

Existing neural decoding architectures can be grouped into several broad classes. Early neural decoders used multilayer perceptrons (MLPs) on flattened syndromes~\cite{krastanov2017deep,torlai2017neural,varsamopoulos2017decoding}. Convolutional neural networks (CNNs) exploit spatial locality in syndrome volumes~\cite{varsamopoulos2019comparing,gu2026scalable}, while recurrent models and high-capacity foundation decoders emphasize the temporal structure generated by repeated syndrome extraction~\cite{baireuther2018machine,varbanov2025neural,bausch2024learning,senior2025scalable}. Graph-based approaches either process detector or spacetime graphs directly~\cite{lange2025data}, or unroll check-to-variable message passing on the decoding graph, as in neural belief propagation~\cite{liu2019neural,maan2025machine,hu2025efficient}.

More recently, attention-based and Transformer decoders have been introduced for neural decoding. These models treat detectors as tokens, giving the decoder a global receptive field over the syndrome history~\cite{bausch2024learning,senior2025scalable}. For example, Transformer-QEC studies cross-distance fine-tuning by re-adapting positional encodings~\cite{wang2023transformerqec}.

Beyond these mainstream architectures, further variants include hypergraph-based, generative and uncertainty-aware neural decoders~\cite{bhave2025hypernq,cao2023qecgpt,mi2025toward}.

We remark that \ntu\ is backbone-agnostic. It can be applied to different neural architectures, provided that their learnable parameters are tied to scale-invariant detector relations, such as the relative-shift set $\mathcal{M}(x,y,t;d)$ in Eq.~(\ref{eq:neighbor}), rather than to absolute detector indices.

\smallskip
\noindent\textbf{Foundation decoders.} A foundation decoder refers to a \textit{high-capacity} neural decoder designed for the \textit{large-scale} regime of QEC, where the goal is to learn reusable decoding representations that remain effective at high code distances and can support downstream adaptation across related decoding tasks~\cite{bausch2024learning,senior2025scalable}. Different neural decoders, foundation decoders emphasize the scale and generality of the learned decoder. That is, the model is trained to capture broadly reusable syndrome structure and to operate in regimes where conventional decoders may become inaccurate or computationally constrained.

Technically, a foundation decoder remains a neural decoder in the sense above, i.e., a training set $\mathcal{T}_d$ is first constructed, followed by optimizing $f(S_d;\bm{\theta})$. In addition, it can be built under either logical error classification in Eq.(\ref{eq:lec}) or physical error inference in Eq.(\ref{eq:pei}), and it can use different neural backbones. What distinguishes it from a general neural decoder is the intended role of its parameters: they should provide a reusable decoding prior for a family of related code instances. This requirement creates a computational challenge. As the code distance grows, the detector volume $m_d$ increases and the model must acquire higher-capacity representations, making training from scratch at the largest distance prohibitively expensive. 

Existing foundation decoders address this challenge through curriculum learning, training first at small distances and progressively enlarging the code~\cite{bengio2009curriculum,senior2025scalable}. However, such procedures still require substantial distance-by-distance adaptation. \ntu\ instead constructs foundation decoders by transfer: relation-tied parameters are reused across distances, so that decoding knowledge learned at smaller distances can be amortized rather than relearned from scratch at every scale.

\subsection{Transfer learning}\label{subsec:pre_transfer}
\smallskip
\noindent\textbf{Parameter transfer.}
Transfer learning reuses knowledge from a source task to improve optimization or generalization on a related target task~\cite{pan2010survey}. In neural networks, this is commonly implemented through parameter initialization. A source model is first optimized to obtain parameters $\bm{\theta}^*$, and a target model is then initialized from $\bm{\theta}^*$, or from a shape-compatible subset of these parameters, instead of random weights. The target model is subsequently fine-tuned on the target task.

This procedure can accelerate convergence and improve data efficiency even when training from scratch could eventually reach comparable accuracy given sufficient data and computation~\cite{yosinski2014how,he2019rethinking}. This effect is often attributed to hierarchical feature reuse: early and intermediate layers encode broadly reusable structure, whereas later layers are more specialized to the target objective and label space~\cite{zamir2018taskonomy}.

Effective parameter transfer is not automatic. It requires the transferred parameters to retain compatible roles in the source and target models. When the input structure, output space or data distribution changes, directly reusing all parameters can lead to negative transfer, in which source-specific features slow down optimization or degrade target performance. In this regard, a central challenge in transfer learning is to decide which parameters should be reused, frozen, reinitialized or fine-tuned, and how to ensure that the reused parameters represent features that remain meaningful for the target task.

\smallskip
\noindent\textbf{Cross-distance transfer for QEC.}
For QEC decoding, \ntu\ casts cross-distance scaling as exactly this source-optimization, parameter-reuse, and target-adaptation pipeline, instantiated on the datasets of the main text. Concretely, training the source decoder $f(S_d;\bm{\theta})$ on $\mathcal{T}_d=\{(S_d^{(i)},\bm{y}_d^{(i)})\}_{i=1}^{N_d}$ at a small, tractable distance $d$ and transferring it to a larger code $d'>d$ amounts to
\begin{equation}
\bm{\theta}^{*}=\arg\min_{\bm{\theta}}\mathcal{L}(\mathcal{T}_{d};\bm{\theta}),
\qquad
\bm{\theta}\leftarrow\bm{\theta}^{*},
\qquad
\bm{\theta}_{d'}^{*}=\arg\min_{\bm{\theta}}\mathcal{L}(\mathcal{T}_{d'};\bm{\theta}),
\label{eq:transfer}
\end{equation}
where $\mathcal{L}$ is the training loss of Eq.~(\ref{eq:training_loss}) and $\mathcal{T}_{d'}=\{(S_{d'}^{(i)},\bm{y}_{d'}^{(i)})\}_{i=1}^{N_{d'}}$. The middle step $\bm{\theta}\leftarrow\bm{\theta}^{*}$ is the transfer itself, namely that the target optimization at $d'$ starts from the source optimum $\bm{\theta}^{*}$ rather than from random weights, with $d'$ a larger lattice for surface codes or larger cyclic blocks $\ell,m$ for BB codes. Only the relation-tied parameters are reused, while the distance-dependent buffers are regenerated for the target code, as detailed in SI~\ref{apx:NTU}.

\smallskip
\noindent\textbf{Transfer with architectural scaling.} Parameter transfer does not require the source and target neural networks to have identical architectures. When the target task requires greater capacity, the source network can be embedded into a larger model through function-preserving network expansion, such as Net2Net transformations~\cite{chen2015net2net}. These transformations widen layers or insert additional layers while initializing the enlarged model to preserve, or minimally perturb, the source function. The newly introduced parameters then provide additional capacity during fine-tuning. 

In this work, however, we keep the hidden dimension and depth fixed across code distances. This fixed-capacity design is a deliberate experimental control, not a requirement of \ntu, and allows us to isolate the effect of relation-tied parameter transfer from the effect of increasing model size.

\section{Details of the \ntu\ Framework}\label{apx:NTU}
Here we formalize the \ntu\ framework outlined in the main text and introduce the definitions used by both decoder instantiations. We first define a scale-invariant code family and make precise the neighbourhood relations in Eq.~(\ref{eq:neighbor}). We then specify the code classes covered by the framework and state a generator-preservation condition for algebraic code families. Finally, we connect this code-level invariance to parameter sharing in neural decoders, describe architecture-agnostic ways to realize it, and present the cross-distance transfer-adaptation protocol. 

\smallskip
\noindent\textbf{Scale-invariant QEC code families.}  The starting point of \ntu\ is a scalable QEC code family whose decoding problem has the same local relational structure at different code distances or block sizes. Informally, a QEC code family is \textit{scale invariant} for decoding if increasing the size of the code only creates more instances of the same detector–detector and detector–qubit relations, rather than changing the rules that define those relations. In other words, the detector and qubit sets may grow with the instance index $d$, but the local relation template used to construct their neighborhoods should remain the same.

We make this notion operational as follows. Let $\mathcal{D}_d$ and $\mathcal{Q}_d$ denote the detector and qubit sets of the code instance indexed by $d$. Each detector $v\in\mathcal{D}_d$ is assigned an algebraic or geometric coordinate $v=x^i y^j t^r$, where $x$ and $y$ encode repeated spatial or algebraic directions and $t$ indexes the syndrome-extraction round. The local decoding environment of $v$ is specified by a finite set of relative shifts $\mathcal{M}(x,y,t;d)$, i.e., $\mathcal{N}(v;d)=v\cdot\mathcal{M}(x,y,t;d)$ as defined in Eq.~(\ref{eq:neighbor}). The QEC code family is \textit{scale invariant} when there exists a common relative-shift template such that, for each distance $d$, $\mathcal{M}(x,y,t;d)$ is obtained by applying this same template to the finite detector set $\mathcal{D}_d$. More specifically, the dependence on $d$ in $\mathcal{M}(x,y,t;d)$ only records finite-instance effects, i.e., detector types, missing neighbors at boundaries, and validity constraints imposed by the finite code geometry. If a shifted coordinate lies outside $\mathcal{D}_d$ or corresponds to an invalid detector type, that neighbor is removed by a validity mask. 

This definition also gives a practical way to identify scale invariance. One first chooses a coordinate system for the detector set, and then checks whether the neighborhood of each detector can be generated from a finite set of relative shifts that is reused across distances. For the rotated surface code, an interior detector at spacetime coordinate $(i,j,r)$ has the same relative pattern of nearby detectors and data-qubit fault mechanisms as an interior detector of the same type in a larger-distance code. Increasing $d$ therefore does not change the local decoding rule; it only places more copies of this rule on a larger spacetime syndrome volume. Boundary detectors are handled by truncating the same relative-shift template to the valid detector set $\mathcal{D}_d$. This repeated local detector environment is the scale invariance of the surface-code decoding problem, as illustrated in Fig.~\ref{fig:scheme}a.

\smallskip
\noindent\textbf{Scope and transferability of scale-invariant families.} The scale-invariant condition is a structural criterion rather than a restriction to a particular named code family. As a result, \ntu\ can support a broad range of QEC code families whose scalable construction preserves a finite typed alphabet of decoder-relevant relations across sizes. Equivalently, after choosing suitable detector coordinates, the local detector neighbourhoods $\mathcal{N}(v;d)$ can be generated from a common relative-shift template, denoted by $\mathcal{M}(x,y,t;d)$ after restriction to the finite instance, up to detector types and boundary truncations.

This criterion covers two broad types of scalable structure. One route to this condition is \textit{geometric repetition}. Regular planar topological code families, such as the rotated surface-code family studied here~\cite{dennis2002topological,fowler2012surface}, preserve their local relation alphabet because the same plaquette or cell pattern is repeated as the code distance increases. Related examples include color-code~\cite{bombin2006topological}, subsystem-code and Floquet-code families~\cite{hastings2021dynamically,davydova2023floquet}, whenever their stabilizer or measurement schedules preserve a finite set of detector-relation types across sizes.

A second route is \textit{algebraic repetition}. Here the reusable relation alphabet is generated not by a repeated geometric cell, but by a fixed algebraic generator template. BB codes described in SI~\ref{subsec:pre_bb} provide such an example~\cite{bravyi2024high}, i.e., increasing the cyclic dimensions while keeping the same polynomial relation pattern preserves the local Tanner-graph alphabet. In this case, generator preservation is the concrete condition that makes relation-tied neural parameters directly reusable across sizes. More generally, lifted-product, hypergraph-product and generalized lifted-product codes can fall within the same scale-invariant class when their scalable construction rules preserve such reusable algebraic relation types~\cite{panteleev2021quantum,breuckmann2021quantum,panteleev2022asymptotically}.

The above distinction also clarifies what lies outside the direct scope of \ntu. A sequence of code instances is not scale invariant if its local detector neighborhoods cannot be generated from a common relative-shift template, either geometrically or algebraically. For example, independently optimized code instances may use different plaquette layouts, measurement schedules or algebraic generators at different sizes. Such choices may improve the code distance or the figure of merit $kd^2/n$~\cite{bravyi2010tradeoffs,bravyi2024high}, but they can also change the relation alphabet on which the neural decoder was trained. While this does not rule out transfer in general, it weakens direct relation-tied parameter reuse and may require reinitializing, remapping or adapting relation-specific embeddings and message functions.

\smallskip
\noindent\textbf{Relation-compatible neural representations and transfer.} We now explain how \ntu\ turns code-level scale invariance into neural parameter reuse. The key requirement is \emph{relation compatibility}: learnable parameters should be tied to reusable decoder-relevant relations, rather than to absolute detector or qubit identities. In this way, these relations encode the scale-invariant structure of a given code family. As a result, the neural decoder represents syndromes $S_d$ through relation types that retain the same meaning across code distances, allowing the corresponding parameters to be transferred. In the notation of Eq.~(\ref{eq:aggregation}), this corresponds to constructing local representations such as $\mathbf{h}(v)$ from relation-indexed weights. As the code is scaled, the detector and qubit sets expand, but the meaning of each relation label is preserved, allowing the learned local perception rule to be reused on the larger instance.

The requirement of relation compatibility is independent of the particular neural backbone, but different architectures realize it through different parameter-sharing mechanisms. To be concrete, convolutional models can implement relation compatibility through shared local stencils associated with fixed relation types~\cite{varsamopoulos2019comparing,gu2026scalable}.  Graph neural networks and neural belief-propagation models implement it through shared message and update functions over node or edge relations~\cite{liu2019neural,maan2025machine}. Transformer-based decoders require additional care, because absolute sequence-position embeddings~\cite{vaswani2017attention} are tied to a particular detector ordering and do not by themselves preserve cross-distance relations. With such relation-preserving representations, the same attention blocks can operate on detector sets of different sizes while retaining access to the local relations learned at smaller distances.

Given a scale-invariant code family and a relation-compatible neural architecture, \ntu\ proceeds cross-distance transfer by separating reusable learnable parameters from distance-dependent auxiliary objects, which provides a structurally aligned initialization that avoids relearning the same decoder-relevant relations from random initialization. Following SI~\ref{subsec:pre_transfer}, the source neural decoder is first trained at a smaller distance $d$, where syndrome generation and optimization are more tractable. Its relation-tied parameters, including embedding tables, local aggregation weights, message functions, recurrent layers, attention blocks and other shape-compatible modules, are then used to initialize the target decoder at a larger distance $d’>d$, following the source-to-target initialization in Eq.~(\ref{eq:transfer}). By contrast, non-learned objects that depend explicitly on the target size, such as coordinate maps, gather indices, sparse edge tensors, attention masks and boundary-validity masks, are regenerated from the target code description. The optimizer state is reset, and the target model is fine-tuned to adapt to boundary statistics, changed detector counts, global aggregation behavior and possible shifts in the syndrome distribution.

\section{Implementation details of \ntu-Transformer}\label{apx:transformer} 

\ntu-Transformer is the Transformer instantiation of the \ntu\ framework in SI~\ref{apx:NTU}, specialized to the logical error classification paradigm of Eq.~(\ref{eq:lec}). Here we describe its implementation details. 

\begin{figure}
    \centering
    \includegraphics[width=\linewidth]{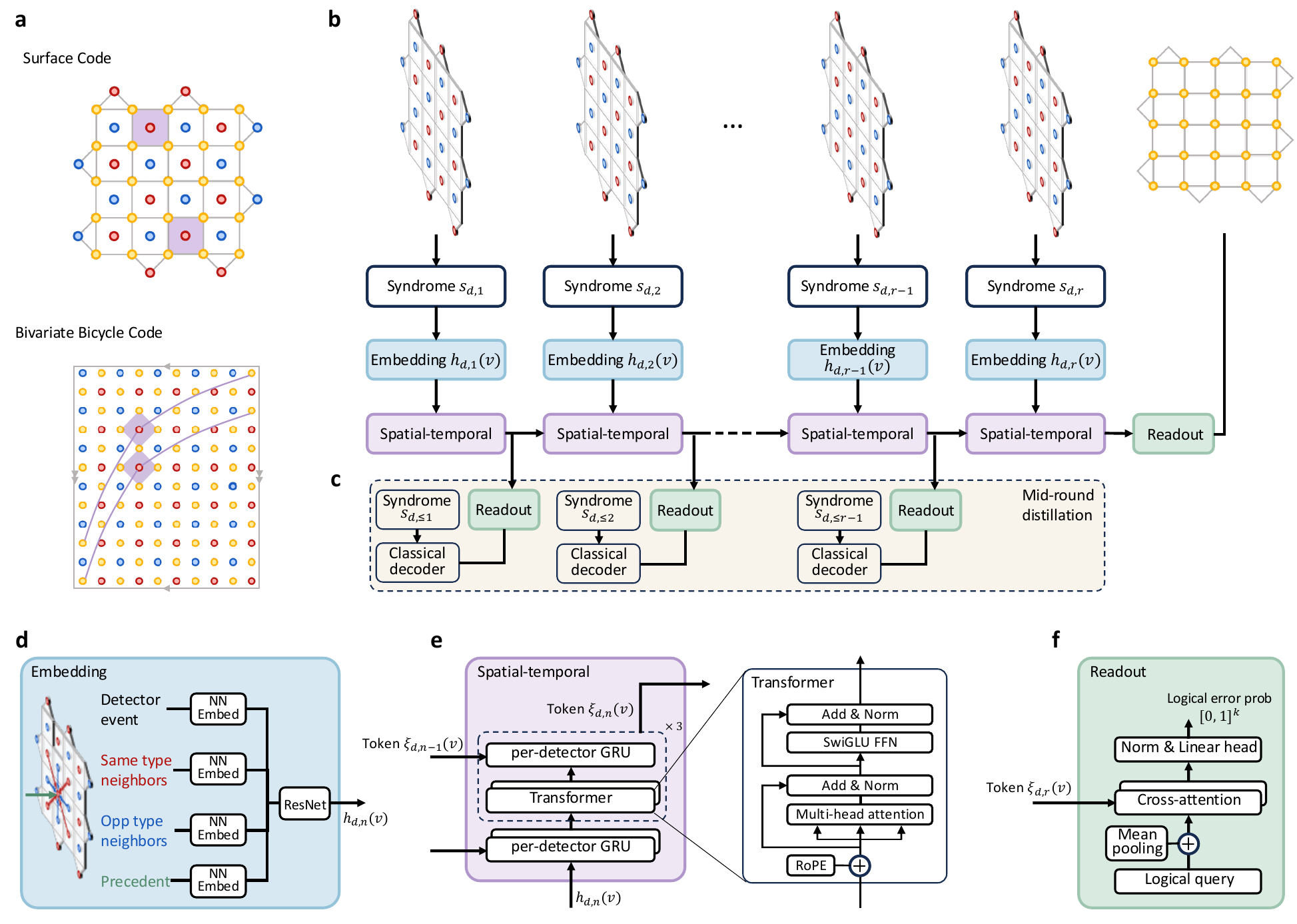}
\caption{
\textbf{Schematic illustration of \ntu-Transformer using the surface code as an example.}
\textbf{a,} Visualization of the scale-invariant detector neighborhood used by the embedding and positional modules. Both codes have data qubits in yellow. X and Z stabilizers are shown in red and blue respectively. Notice that the BB codes are cyclic codes with periodic boundaries.
\textbf{b,} Overview of the \ntu-Transformer pipeline over a detector-event history. At each round, the syndrome vector is converted into topology-driven detector embeddings, processed by the spatiotemporal backbone, and finally aggregated by the readout module to predict the logical error probability.
\textbf{c,} Mid-round process supervision. Prefix syndromes are decoded by a classical teacher to produce intermediate pseudo-labels, which provide auxiliary supervision during training while using only observable detector prefixes.
\textbf{d--f,} Implementation details of the main network blocks. 
\textbf{d,} The embedding block combines detector events with same-type, opposite-type, and temporal relation features before a residual stem projection. This relation-indexed construction is shared by surface and BB codes once the corresponding neighborhood relations are specified.
\textbf{e,} The spatiotemporal block alternates recurrent detector-wise updates with spatial Transformer layers. The Transformer layer applies geometry-aware RoPE to attention queries and keys. For BB codes, the coordinates and offsets used by RoPE are defined by the polynomial relation system to preserve transferability across block sizes.
\textbf{f,} The readout block uses logical-query cross-attention to aggregate variable-size detector representations into fixed-size logical error probabilities.
}\label{fig:apx-transformer}
\end{figure}

The pipeline of \ntu-Transformer is summarized in SI Fig.~\ref{fig:apx-transformer} and consists of \textit{three stages}. First, \textit{topology-driven input embeddings} convert the multi-round detector history $S_d\in\mathbb{F}_2^{m_d}$ into detector tokens $\{\xi_{d,r}(v):v\in\mathcal{D}_d\}$ for subsequent recurrent and attention updates. For both surface and BB codes, the local component of $\xi_{d,r}(v)$ is constructed from the binary syndrome pattern in the invariant neighbourhood $\mathcal{N}(v;d)$. Its features are therefore indexed by reusable code relations rather than absolute detector positions, following the relation-indexed representation $\mathbf{h}(v)$ in Eq.~(\ref{eq:aggregation}).

In the second stage, a factorized \textit{spatiotemporal backbone} updates the tokens $\{\xi_{d,r}(v)\}$ by combining per-detector recurrent updates along the syndrome-extraction rounds with spatial Transformer layers equipped with \textit{geometry-aware rotary positional encodings}. In the third stage, a \textit{logical-query cross-attention head} aggregates the variable-size token set into the fixed-size logical prediction $\hat{\bm{y}}_d\in\mathbb{F}_2^k$ in Eq.~(\ref{eq:lec}).

The remainder of this section is organized as follows. In SI~\ref{subsec:transformer_arch}, we describe the spatiotemporal backbone together with the three code-aware components that make this pipeline compatible with the scale-invariant structure of the code family: scale-invariant input embeddings, geometry-aware rotary positional encodings for relation-preserving global attention, and logical-query cross-attention for fixed-size logical readout. In SI~\ref{subsec:transformer_train}, we describe the mid-round distillation signal used during training, which provides additional supervision on detector-prefix inputs and helps mitigate the cold-start plateau.

\subsection{Implementation details of core components in \ntu-Transformer}\label{subsec:transformer_arch}

\smallskip
\noindent\textbf{Topology-driven input embedding.}
\ntu-Transformer is designed to operate on binary detector events rather than continuous analog readouts. This distinction is important for the first stage of representation learning. In hardware-specific neural decoders with access to soft measurement information, the input already contains continuous confidence values, and a direct linear or multilayer-perceptron projection can extract dense low-level features~\cite{bausch2024learning,lacroix2025scaling}. In our setting, the available input is a sparse binary detector-event history $S_d$. At the low physical error regime, most detector entries are zero, while the informative nonzero events are rare and highly localized. Our empirical studies observe that directly projecting this sparse binary tensor with a generic dense layer or convolution is very challenging, in which the   early optimization is unstable. This is caused by the fact that most input channels are repeatedly inactive and gradients are concentrated on a small number of rare event locations.

To address this challenge, \ntu-Transformer uses topology-driven lookup embeddings, as illustrated in SI Fig.~\ref{fig:apx-transformer}d. Instead of assigning embeddings to absolute detector indices, the input embedding module builds a dense representation from the binary pattern observed in the invariant neighborhood $\mathcal{N}(v;d)$, drawing from a fixed set of lookup entries indexed by the relation set $\mathcal{M}(x,y,t;d)$ of the code family~\cite{sadhukhan2026stem}. Concretely, the embedding is the relation-indexed lookup, i.e.,
\begin{equation}
\mathbf{h}(v)=\sum_{r\in\mathcal{M}(x,y,t;d)}\mathbf{E}_r\big[S_d(v\cdot r)\big],
\label{eq:embed}
\end{equation}
where $\mathbf{E}_r[\cdot]$ is a learnable table indexed by the binary syndrome value at the relation-$r$ neighbor, a discrete realization of the aggregation in Eq.~(\ref{eq:aggregation}). When the code distance $d$ increases, the number of detectors grows accordingly, whereas the relation set $\mathcal{M}(x,y,t;d)$ and its tables $\mathbf{E}_r[\cdot]$ remain unchanged, so the same embedding parameters can be reused across distances.

We now specify how the relation set $\mathcal{M}(x,y,t;d)$ that indexes the embedding in Eq.~(\ref{eq:embed}) is instantiated for each code family. For rotated surface codes, the relation system is local and geometric. The embedding in Eq.~(\ref{eq:embed}) uses the binary value of a detector together with nearby same-type, cross-type, and temporal neighbors defined by the plaquette structure. For BB codes, there is no faithful planar nearest-neighbor geometry because the stabilizer connectivity is non-local. We therefore define the embedding neighborhood algebraically from the polynomial-induced stabilizer-intersection structure, as illustrated in SI Fig.~\ref{fig:apx-transformer}a. In particular, the same-type and cross-type detector relations are read off from the generating polynomials $A$ and $B$ of Eq.~(\ref{eq:bb_matrix}). That is, two stabilizer checks share data qubits precisely when the corresponding blocks of $H_X=[A\ B]$ and $H_Z=[B^{T}\ A^{T}]$ have overlapping support, and the relative shift connecting them is the group-ring monomial that links their supports. Here the transpose $(\cdot)^{T}$ appearing in $H_Z$ is the group-ring involution $x\mapsto x^{-1},\,y\mapsto y^{-1}$, which we apply consistently so that each relation is indexed by the physically correct cyclic shift. This avoids heuristic coordinate metrics, such as torus Euclidean distance, which may assign the same embedding label to detector pairs that do not represent the same physical relation. Under this construction, the surface-code and BB-code embeddings follow the same principle: local input features are indexed by invariant code relations.

\smallskip
\noindent\textbf{Spatiotemporal perception backbone.}
\ntu-Transformer processes the embedded detector-event history $S_d$ with a factorized spatiotemporal backbone. The temporal component propagates information across repeated QEC rounds, while the spatial or algebraic component models correlations among detectors within a single round. \ntu-Transformer implements this structure by interleaving gated recurrent layers~\cite{cho2014learning} with spatial Transformer blocks~\cite{vaswani2017attention}. The recurrent layers update detector representations along the time direction, and the Transformer layers allow all detector tokens within the same round to exchange information through self-attention~\cite{vaswani2017attention}.

This design is aligned with the transfer principle of \ntu\ in SI~\ref{apx:NTU}. Each round applies a per-detector recurrent update along time followed by spatial self-attention across detectors, i.e.,
\begin{equation}
\xi_{d,r}(v)=\mathrm{TF}\big(\mathrm{GRU}(\xi_{d,r-1}(v),\,\mathbf{h}_{d,r}(v))\big),
\label{eq:backbone}
\end{equation}
where $\mathbf{h}_{d,r}(v)$ is the round-$r$ embedding of detector $v$ from Eq.~(\ref{eq:embed}), $\mathrm{GRU}(\cdot, \cdot)$ is the gated recurrent update~\cite{cho2014learning} that propagates each detector token along the $R$ rounds, and $\mathrm{TF}(\cdot)$ is a spatial Transformer layer that mixes the tokens $\{\xi_{d,r}(v):v\in\mathcal{D}_d\}$ within round $r$. Because the operations $\mathrm{GRU}$ and $\mathrm{TF}$ share parameters across rounds and detectors, increasing the code distance $d$ changes the number of tokens but not the update rule.

The backbone employed in \ntu-Transformer consists of five gated recurrent layers and six Transformer layers, with Transformer blocks inserted between recurrent stages, as illustrated in SI Fig.~\ref{fig:apx-transformer}e. Each Transformer block uses multi-head self-attention, a SwiGLU feed-forward module, and normalization layers. The main architectural specifications of \ntu-Transformer are summarized in SI Table~\ref{tab:model_spec}.

\smallskip
\noindent\textbf{Geometry-aware rotary positional encoding.} A key challenge in applying Transformers to scalable QEC codes is the absence of a distance-independent positional system. Standard learned positional embeddings~\cite{devlin2018bert} and sequence-indexed sinusoidal embeddings~\cite{vaswani2017attention}, which are broadly used in large language models, are tied to the ordering and length of the detector sequence. However, they are infeasible for QEC  decoding. This is because when the code distance $d$ changes, the detector set $\mathcal{D}_d$ expands and the flattened sequence changes, so the same physical or algebraic relation may receive a different positional representation. This issue is especially severe for non-local qLDPC codes, where the Tanner graph does not provide a natural Euclidean coordinate system for attention.

To overcome this bottleneck, \ntu-Transformer adapts a geometry-aware rotary positional encoding~\cite{su2024roformer} to supply such a coordinate system. In particular, the position assigned to a detector is not its sequence index, but its intrinsic coordinate in the code construction. For rotated surface codes, this coordinate is the physical lattice coordinate of the stabilizer detector. For BB codes, it is the cyclic matrix coordinate induced by the block-circulant polynomial construction. Writing $p_v$ for this coordinate and $R(\cdot)$ for the rotary rotation, the attention logit, i.e., the pre-softmax score between detectors $u$ and $v$, becomes
\begin{equation}
a_{uv}=\big[R(p_u)\mathbf{q}_u\big]^{\top}\big[R(p_v)\mathbf{k}_v\big]=\mathbf{q}_u^{\!\top}R(p_v-p_u)\,\mathbf{k}_v,
\label{eq:rope}
\end{equation}
where $\mathbf{q}_u$ and $\mathbf{k}_v$ are the query and key vectors of detectors $u$ and $v$, $R(p)$ is the block-diagonal rotary rotation matrix associated with coordinate $p$, and $a_{uv}$ is the resulting attention logit. The second equality uses the rotary identity $R(p_u)^{\top}R(p_v)=R(p_v-p_u)$, so $a_{uv}$ depends only on the displacement $p_v-p_u$. Since a fixed relation $v^{-1}u\in\mathcal{M}(x,y,t;d)$ gives a fixed displacement, the same local relation produces the same relative phase at any code size.

\smallskip
\noindent\textit{Remark}. We deliberately do not normalize these coordinates by the global lattice size or torus dimensions. Although normalizing positional indices by the sequence length is sometimes adopted to encourage length generalization in Transformers~\cite{su2024roformer}, here it would make the same local displacement appear smaller as the code grows, changing the relative phase associated with an invariant relation. The rotary encoding is therefore defined on unnormalized code coordinates with a code-size-independent frequency basis. For surface codes, this recovers the usual physical geometric intuition. For BB codes, it provides an algebraic coordinate system that makes polynomial shifts visible to attention, enabling Transformer layers to represent relative structure even in non-local qLDPC geometries.

\smallskip
\noindent\textbf{Logical readout by cross-attention.} For logical error classification in Eq.~(\ref{eq:lec}), the decoder must map the variable-size detector representation to the fixed-size prediction $\hat{\bm{y}}_d$ over the $k$ logical observables. A flattened classification head would be tied to a particular code distance and cannot be reused after scaling. Simple length-independent pooling operations, such as mean or max pooling~\cite{lee2019set}, avoid this shape mismatch but can discard structured parity information that is needed to resolve logically inequivalent error classes.

As shown in SI Fig.~\ref{fig:apx-transformer}f, \ntu-Transformer instead uses a cross-attention readout. Mathematically, a set of $k$ learnable logical query tokens $Q=\{\mathbf{q}^{\mathrm{log}}_1,\dots,\mathbf{q}^{\mathrm{log}}_k\}$, one for each of the $k$ logical observables of the $[\![n,k,d]\!]$ code and distinct from the per-detector attention queries $\mathbf{q}_u$ in Eq.~(\ref{eq:rope}), attends to the final-round detector tokens,
\begin{equation}
\hat{\bm{y}}_d=\mathrm{CrossAttn}\big(Q,\,\{\xi_{d,R}(v):v\in\mathcal{D}_d\}\big),
\label{eq:readout}
\end{equation}
producing the $k$-dimensional prediction $\hat{\bm{y}}_d$ of Eq.~(\ref{eq:lec}). This mechanism functions as a trainable, geometry-aware pooling operation. It is compatible with a detector set $\mathcal{D}_d$ of variable size $m_d$ while still allowing the logical queries to aggregate distributed evidence across the full syndrome history. The readout is shape-compatible across distances. However, because the number of detector tokens changes with code distance $d$, the statistics of attention logits and aggregated features also change. For this reason, the readout is transferred together with the backbone but is further recalibrated during target-distance fine-tuning.

\subsection{Mid-round supervision via classical distillation}\label{subsec:transformer_train}

Training a logical classification decoder only from the final logical label can create a severe optimization bottleneck as the number of rounds scale polynomially with the code distance $d$. At large distances, the syndrome history is long, logical failures are sparse, and the final label provides weak supervision to early recurrent states. In practice, this can cause a randomly initialized model to remain near a constant-output solution during the early stage of training.

To provide denser training signals, \ntu-Transformer uses mid-round supervision via classical distillation, as illustrated in SI Fig.~\ref{fig:apx-transformer}c. More specifically, at an intermediate round, \ntu-Transformer retains the observed detector events up to that round and mask the future detector events. A classical decoder is then applied to this truncated detector history to generate an intermediate pseudo-label. The neural decoder is trained on a combination of the final logical label and these intermediate pseudo-labels, i.e.,
\begin{equation}
\mathcal{L}=\ell(\hat{\bm{y}}_d,\bm{y}_d)+\beta(s)\sum_{r=1}^{R}w_r\,\ell\big(\hat{\bm{y}}_d^{(r)},\tilde{\bm{y}}_d^{(r)}\big),
\label{eq:distill}
\end{equation}
where $\ell$ is the per-sample loss of Eq.~(\ref{eq:training_loss}), $\hat{\bm{y}}_d^{(r)}$ is the prediction from the round-$r$ detector prefix, $\tilde{\bm{y}}_d^{(r)}$ is the classical-teacher pseudo-label on that prefix, $w_r$ are time-progressive weights, and $\beta(s)$ is the distillation coefficient annealed over training step $s$.

This construction is different from fake-end supervision based on simulator-internal information~\cite{senior2025scalable}. Simulator-derived intermediate observables (e.g. Stim-based simulation) can be useful for training, but they are not available from real quantum devices. Our pseudo-labels require only the experimentally accessible detector-event prefix and an external classical decoder. Therefore, the same supervision strategy can in principle be applied to real-device syndrome streams, provided that a suitable classical teacher decoder is available.

The distillation weight is large during the early stage of training, where it supplies stable intermediate gradients. The weight is then reduced as training proceeds, so that the network is not forced to imitate the teacher indefinitely. This annealing is important because the classical teacher may be suboptimal under correlated circuit-level noise or highly degenerate error configurations. Once the auxiliary supervision is weakened, the neural decoder continues optimizing against the final logical labels and can learn corrections beyond the teacher's decision rule.

\section{Implementation details of \ntu-Neural-BP}\label{apx:neural-BP}
 
\begin{figure}
    \centering
    \includegraphics[width=\linewidth]{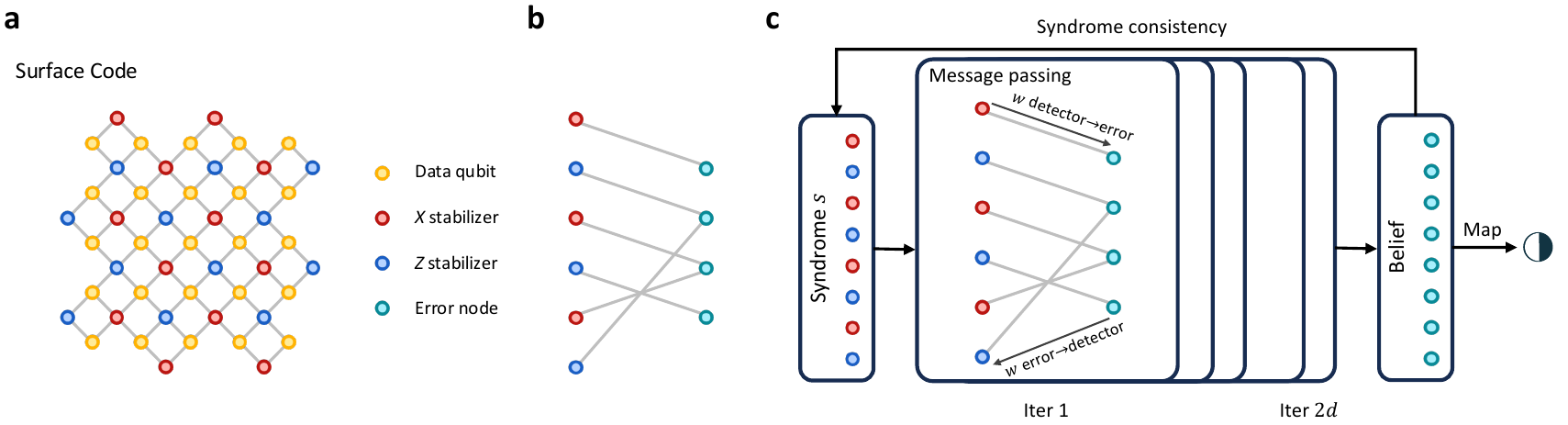}
    \caption{\textbf{Schematic illustration of \ntu-Neural-BP using the rotated surface code as an example.}
\textbf{a,} Local planar structure of the rotated surface code, showing data qubits and the two stabilizer types.
\textbf{b,} Detector-error bipartite graph used by \ntu-Neural-BP. Red and blue nodes denote the two detector types, and green nodes denote error variables defined by the detector error model rather than physical data qubits themselves. In the surface-code circuit-level noise model used here, these error variables arise from four elementary noise channels in Stim: \texttt{after\_clifford\_depolarization}, \texttt{after\_reset\_flip}, \texttt{before\_round\_data\_depolarization}, and \texttt{before\_measure\_flip}. Each green node is connected to the detector nodes flipped by the corresponding DEM term.
\textbf{c,} Neural message-passing decoder. The observed syndrome is injected on the detector side, and shared neural message updates are iteratively applied on the detector-error graph for $2d$ iterations. During training, the inferred error-variable beliefs are additionally constrained by syndrome consistency. After the final iteration, the belief on each error node is used to infer the error configuration, which is then mapped through the fixed logical map to predict the logical outcome.}
    \label{fig:apx-neuralBP}
\end{figure}

\ntu-Neural-BP is the message-passing instantiation of the \ntu\ framework of SI~\ref{apx:NTU}, specialized to the physical error inference paradigm of Eq.~(\ref{eq:pei}). Its pipeline is summarized in SI Fig.~\ref{fig:apx-neuralBP}. It runs directly on the bipartite Tanner graph of SI~\ref{subsec:pre_bb}, whose check nodes are the detectors carrying the syndrome $S_d$ and whose variable nodes are the $M$ fault mechanisms. The shared neural message updates are unrolled for $2d$ iterations, after which the variable beliefs give the recovery $\hat{\bm{e}}\in\mathbb{F}_2^{M}$, mapped through the fixed logical map to $\hat{\bm{y}}_d=L\hat{\bm{e}}$ as in Eq.~(\ref{eq:pei}). In contrast to \ntu-Transformer, which builds transferable coordinates so that global attention sees the same relation at every distance, \ntu-Neural-BP uses the sparse decoding graph itself as the scaffold and shares only local update rules, so its transfer involves no learned positional representation.

The core components of \ntu-Neural-BP are a relation-tied neural message passer that transfers across graph sizes, recurrent gated stabilization for the short-cycle and degeneracy pathologies of belief propagation (BP) on qLDPC graphs, and a syndrome-consistency-regularized inference objective. In the remainder of this section, we first detail these core components in SI~\ref{subsec:neuralbp_arch}. After that, we exhibit the training objective of \ntu-Neural-BP in SI~\ref{subsec:neuralbp_train}.

\subsection{Implementation details of core components in \ntu-Neural-BP}\label{subsec:neuralbp_arch}

\smallskip
\noindent\textbf{Graph-induced relation representation.}
\ntu-Neural-BP provides a message-passing instantiation of \ntu. Instead of constructing detector tokens and positional coordinates as in \ntu-Transformer, it operates directly on the sparse decoding graph associated with the code, as illustrated in SI Fig.~\ref{fig:apx-neuralBP}. In particular, the graph contains detector nodes, error-variable nodes, and edges, indicating which error variables contribute to which detector events. In a code-level formulation, these error variables may correspond to data-qubit error components. In the DEM formulation used for the BB-code experiments, they correspond to elementary error mechanisms, each of which flips a specified set of detectors and logical observables. 

This graph-induced representation plays the same conceptual role as the topology-driven embedding in \ntu-Transformer. It exposes the invariant relation structure of the code to the neural network while avoiding any dependence on absolute detector or variable indices. The graph tensors, including the check matrix $H$ and logical-observable map $L$ of SI~\ref{subsec:pre_bb} and the per-variable priors $p_j$, are generated from the target code instance, whereas the trainable parameters are attached to the shared message and update functions. Mathematically, each error variable $j$ enters message passing through its prior as a log-likelihood ratio
\begin{equation}
\gamma_j=\log\frac{1-p_j}{p_j},
\label{eq:bp_prior}
\end{equation}
and the observed syndrome $S_d$ is injected on the check (detector) nodes. Thus, when the code is scaled, the graph becomes larger, but the local computational rule applied on the graph remains reusable. 

\smallskip
\noindent\textbf{Relation-tied neural message passing.} 
As depicted in the neural message-passing decoder of SI Fig.~\ref{fig:apx-neuralBP}c, \ntu-Neural-BP follows the information flow of belief propagation while replacing fixed analytic updates with trainable neural updates. Messages are exchanged between detector nodes and error-variable nodes along the sparse graph. The same message-generation and node-update functions are shared across graph locations, so the model learns how to process a type of local relation rather than how to process a particular edge index. Concretely, at iteration $t$, the variable-to-check messages are produced by a shared neural function, i.e.,
\begin{equation}
m_{j\to c}^{(t)}=g_{\bm{\theta}}\Big(\gamma_j,\ \textstyle\sum_{c'\ni j,\,c'\neq c} m_{c'\to j}^{(t-1)}\Big),
\label{eq:bp_message}
\end{equation}
where $j$ indexes an error-variable node and $c$ a check (detector) node, $m_{j\to c}^{(t)}$ is the message sent from variable $j$ to check $c$ at iteration $t$, and $\gamma_j$ is the prior log-likelihood ratio of Eq.~(\ref{eq:bp_prior}). The sum runs over all checks $c'$ incident to variable $j$ except the recipient $c$ (denoted $c'\ni j,\,c'\neq c$), aggregating the incoming check-to-variable messages $m_{c'\to j}^{(t-1)}$ from the previous iteration; excluding the recipient $c$ is the standard extrinsic-information rule of belief propagation. The shared neural function $g_{\bm{\theta}}$ is applied identically at every variable node, and an analogous shared function generates the check-to-variable messages. Because $g_{\bm{\theta}}$ is tied to relation type rather than to an edge index, it transfers unchanged to a larger graph.

This parameter sharing is the message-passing analogue of the relation-tied representation used by \ntu-Transformer. In \ntu-Transformer, the embedding and positional encoding are designed so that attention sees the same relative relation at different distances. In \ntu-Neural-BP, the graph itself specifies the allowed relations, and the neural update rule is applied uniformly across them. The resulting decoder is therefore naturally compatible with cross-distance transfer as long as the scaled code family preserves the same decoding-graph relation structure. 

\smallskip
\noindent\textbf{Stabilized neural BP updates.} This component stabilizes the unrolled message passing of SI Fig.~\ref{fig:apx-neuralBP}c and comprises three ingredients: multilayer perceptrons (MLPs) that transform the messages, recurrent gated units that maintain a per-node state across iterations, and normalization with bounded nonlinearities that control the message scale. Stabilization is necessary because qLDPC decoding graphs contain many short cycles and degenerate error configurations, which can make conventional belief propagation oscillate or become trapped in unreliable fixed points. \ntu-Neural-BP mitigates these pathologies by parameterizing the message-passing dynamics with these neural modules: the MLPs transform messages, while the recurrent gated updates maintain detector and error-variable states across iterations. The recurrent state allows the model to regulate repeated information circulating through short loops, rather than applying a memoryless BP update at every step. Each node maintains a recurrent state updated by a gated unit~\cite{cho2014learning},
\begin{equation}
h_j^{(t)}=\mathrm{GRU}\big(h_j^{(t-1)},\, m_j^{(t)}\big),
\label{eq:bp_state}
\end{equation}
where $h_j^{(t)}$ is the recurrent state of node $j$ at iteration $t$ and $m_j^{(t)}$ aggregates the incoming messages of Eq.~(\ref{eq:bp_message}). The memory in $h_j^{(t)}$ damps the oscillations that short cycles induce in memoryless BP.

Finally, normalization layers and bounded nonlinearities control the scale of the latent messages as the unrolled iteration depth increases; these components leave the decoding graph unchanged and only make the learned message dynamics easier to optimize. In the BB-code experiments, the resulting \ntu-Neural-BP model has approximately $3\times 10^5$ trainable parameters, with the concrete depth, optimizer settings, and training schedule reported in SI~\ref{apx:exp_details} and the full architecture specifications listed in SI Table~\ref{tab:model_spec}.

\subsection{Training: physical error inference and syndrome consistency}\label{subsec:neuralbp_train}

We first describe how \ntu-Neural-BP is trained at a single code distance through a physical-error-inference objective, and then how the trained model is transferred across graph sizes.

\smallskip
\noindent\textbf{Inference objective.}
At a single code distance, \ntu-Neural-BP is trained in the physical-error-inference paradigm of Eq.~(\ref{eq:pei}). Here, physical error inference means predicting the low-level recovery $\hat{\bm{e}}\in\mathbb{F}_2^{M}$ on the decoding graph before its logical effect $\hat{\bm{y}}_d=L\hat{\bm{e}}$ is determined through the fixed logical map $L$ of SI~\ref{subsec:pre_bb}. For the BB-code implementation, the entries of $\bm{e}$ are the $M$ elementary error mechanisms in the detector-error model. After prediction, the inferred error variables are mapped to detector syndromes and logical observable flips using the corresponding graph and logical maps. The prediction target is sparse at low physical error rates, so the supervised objective is chosen to emphasize rare active error variables and to avoid collapse to the trivial no-error prior. We also include a syndrome-consistency regularization term, giving the objective
\begin{equation}
\mathcal{L}=\mathcal{L}_{\mathrm{focal}}(\hat{\bm{e}},\bm{e})+\lambda_{\mathrm{syn}}\,\big|H\hat{\bm{e}}\oplus S_d\big|,
\label{eq:bp_loss}
\end{equation}
where $\hat{\bm{e}}$ is the predicted error configuration read from the final beliefs, $\bm{e}$ is the true error label, and $\mathcal{L}_{\mathrm{focal}}$ is the focal loss~\cite{lin2017focal}, which down-weights the abundant inactive variables and emphasizes the rare active ones. The second term penalizes syndrome inconsistency with weight $\lambda_{\mathrm{syn}}$: here $\oplus$ denotes bitwise modulo-$2$ addition (XOR), so $H\hat{\bm{e}}\oplus S_d$ is the residual syndrome of the prediction and $\big|H\hat{\bm{e}}\oplus S_d\big|$ its Hamming weight, which vanishes exactly when $H\hat{\bm{e}}=S_d$. This injects the parity-check constraint into the optimization and encourages physically admissible predictions. During evaluation, the predicted $\hat{\bm{e}}$ is mapped to $\hat{\bm{y}}_d=L\hat{\bm{e}}$ as in Eq.~(\ref{eq:pei}), and a failure is counted when $\hat{\bm{y}}_d\neq\bm{y}_d$. 

\smallskip
\noindent\textbf{Transfer across graph sizes.}
Cross-distance transfer in \ntu-Neural-BP separates learnable local update rules from non-learned graph tensors. The source model is first trained on a smaller code instance. Its trainable message functions, recurrent update functions, normalization parameters, and readout components are then loaded into the target model. The sparse graph tensors for the target code, including detector indices, error-variable indices, edge connectivity, priors, and logical maps, are regenerated from the target code description. This transfer mechanism is simpler than that of \ntu-Transformer because no learned absolute positional representation is involved. The graph defines which objects communicate, and the neural parameters define how information is transformed along those relations. Target-side fine-tuning then adapts the reused message-passing dynamics to the changed graph size, iteration depth, and syndrome distribution. 

\smallskip
\noindent\textbf{Extension to other backbones.} The same relation-tied principle extends beyond the two backbones studied here to other graph neural decoders, hypergraph neural decoders, or trainable variants of belief propagation, provided that their parameters are tied to invariant node, edge, or hyperedge relation types rather than to absolute indices.

\begin{table}[t]
\centering
\small
\setlength{\tabcolsep}{4.5pt}
\renewcommand{\arraystretch}{1.12}
\caption{
Model architecture specifications for the two \ntu\ decoder instantiations.
\ntu-Transformer performs logical error classification with relation-aware attention, whereas \ntu-Neural-BP performs physical error inference through relation-tied message passing. Training hyperparameters and optimization protocols are reported in SI~\ref{apx:exp_details}.
}
\label{tab:model_spec}
\begin{tabular}{lll}
\toprule
\textbf{Configuration module} & \textbf{\ntu-Transformer} & \textbf{\ntu-Neural-BP}  \\
\midrule
Decoding paradigm
& Logical error classification
& Physical error inference \\

Input structure
& Binary detector tokens
& Detector-error graph \\

Core dimensions
& $d_{\mathrm{model}}{=}512$, $n_{\mathrm{heads}}{=}8$
& $d_{\mathrm{hidden}}{=}128$ \\

Network depth
& 5 GRU layers, 6 TF layers
& 12/24 BP iterations \\

Feed-forward / MLP size
& SwiGLU, $d_{\mathrm{hidden}}{=}1024$
& Message-update MLPs, $d_{\mathrm{hidden}}{=}128$ \\

Readout mechanism
& 2-layer cross-attention, $d_{\mathrm{FFN}}{=}2048$
& 2-layer MLP with GELU \\

Relation encoding
& Topological embedding + RoPE, $\Theta{=}100$
& Sparse graph connectivity \\

Stabilization layers
& RMSNorm
& LayerNorm, ScaledTanh \\

Total parameters
& $\sim 32$M
& $\sim 0.30$M \\
\bottomrule
\end{tabular}
\end{table}

\section{Experimental details}\label{apx:exp_details}
This section is organized as follows. SI~\ref{subsec:eval_metrics} defines the evaluation metrics, in particular the logical error per round used throughout. SI~\ref{subsec:exp_surface} and SI~\ref{subsec:exp_bb} then give the experimental protocols, training hyperparameters, and classical baselines for the surface-code and BB-code experiments, respectively.

\smallskip
\noindent\textit{Remark}. Unless otherwise stated, neural-decoder training was performed with PyTorch distributed data parallelism on NVIDIA H800 GPUs with 80 GB memory. \ntu-Transformer experiments used bfloat16 mixed precision, whereas \ntu-Neural-BP experiments were run in float32. Online syndrome generation and, when applicable, teacher-label generation were parallelized with CPU workers and streamed to the GPU training processes. The large-distance surface-code stress tests at $d=23$ and $d=25$ used a separate 16-GPU NVIDIA H200 configuration with 140 GB memory per GPU, as described in SI~\ref{apx:results}.

\subsection{Evaluation metrics}\label{subsec:eval_metrics} 
\smallskip
\noindent\textbf{The calculation of logical error per round and per logical qubit.}
To ensure fault tolerance in the time domain, each quantum memory experiment is executed across exactly $d$ consecutive syndrome extraction rounds to form a complete spatiotemporal code block. The total logical error rate of this block is defined as the complement of the decoding accuracy over the $d$-round duration, meaning the temporal depth of the block scales proportionally with the code distance $d$. To facilitate a standardized comparison of decoding performance across varying experiment lengths and structurally distinct code families, we adopt the logical error per round (LER) as our primary evaluation metric~\cite{google2023suppressing,senior2025scalable,google2025quantum}. For architectures encoding multiple logical qubits ($k > 1$), such as BB codes, the metric is strictly quantified as the logical error per round and per logical qubit~\cite{ruiz2025ldpccat}. Unless explicitly stated otherwise, the phrase ``per qubit'' throughout this work always denotes per logical qubit, and a single round refers to one full QEC cycle.
 
Let $P_{\mathrm{block}}(R)$ denote the block-level logical error probability after $R$ rounds of syndrome extraction. For a single logical qubit, we define the logical fidelity as
\begin{equation}
F(R)=1-2P_{\mathrm{block}}(R).
\end{equation}
Assuming independent round-by-round accumulation of logical errors, the fidelity satisfies
\begin{equation}
F(R)=\left(1-2\epsilon_{\mathrm{round}}\right)^R,
\end{equation}
where $\epsilon_{\mathrm{round}}$ is the LER per round. Thus,
\begin{equation}
\epsilon_{\mathrm{round}}
=
\frac{1}{2}
\left[
1-\left(1-2P_{\mathrm{block}}(R)\right)^{1/R}
\right].
\end{equation}
If results at multiple round numbers $R$ are available, we instead fit
\begin{equation}
\log F(R)=\log F_0+R\log\left(1-2\epsilon_{\mathrm{round}}\right),
\end{equation}
and extract $\epsilon_{\mathrm{round}}$ from the fitted slope. This logical-fidelity-based extraction is consistent with the LER definition adopted in AlphaQubit~2 and related QEC studies~\cite{senior2025scalable,google2025quantum,google2023suppressing}.

\smallskip
For a code encoding $k$ logical qubits, we further define the block-level logical error probability per logical qubit as
\begin{equation}
P_{\mathrm{LQ}}(R)=1-\left(1-P_{\mathrm{block}}(R)\right)^{1/k}.
\end{equation}
This normalization converts the total block-level logical error probability into the effective logical error probability of a single logical qubit, and is introduced here to enable fair comparison between code families with different logical encoding rates~\cite{ruiz2025ldpccat}. The corresponding LER per logical qubit per round is
\begin{equation}
\epsilon_{\mathrm{LQ}}
=
\frac{1}{2}
\left[
1-\left(1-2P_{\mathrm{LQ}}(R)\right)^{1/R}
\right].
\end{equation}
Unless otherwise stated, the LER per qubit is referred as to $\epsilon_{\mathrm{LQ}}$. In the low-error regime, $P_{\mathrm{block}}(R)\ll 1$, this reduces to
\begin{equation}
P_{\mathrm{LQ}}(R)\approx \frac{P_{\mathrm{block}}(R)}{k},
\qquad
\epsilon_{\mathrm{LQ}}\approx \frac{P_{\mathrm{block}}(R)}{kR}.
\end{equation}
This approximation is useful for intuition, while all reported values in this study are computed using the exact expressions above.
\smallskip

\smallskip
\noindent\textbf{Computes and training time.}
We profile computational overhead using two complementary metrics. First, we estimate algorithmic training compute from multiply--accumulate operations (MACs). Forward MACs are measured with the \texttt{fvcore} library using the model in \texttt{train()} mode, so that auxiliary branches used for process supervision are included in the profiled computation. Following the standard approximation that a training step costs about three forward passes when accounting for forward and backward propagation~\cite{kaplan2020scaling}, we estimate the training cost per step as 3 times the profiled forward MACs. We then convert MACs to floating-point operations by counting one MAC as two FLOPs. Thus, the reported training compute per step is 6 times the profiled forward MAC count. Unless stated otherwise, computational budgets in this work are reported in FLOPs.

Second, we measure empirical throughput using wall-clock profiling. Because CUDA execution in PyTorch is asynchronous, we place \texttt{torch.cuda.synchronize()} calls around the timed code regions before recording elapsed time. For each training step, we separately measure the online data generation and preprocessing time, the GPU forward--backward computation time, and the distributed data-parallel communication plus optimizer-update time. This decomposition separates model computation from data-generation and synchronization overheads, allowing throughput comparisons across code distances and model configurations to be interpreted more consistently.

\subsection{Experimental setup for surface code}\label{subsec:exp_surface} 
\smallskip
\noindent\textbf{Experimental protocol.}
For the surface-code experiments, we generate rotated surface-code memory circuits with $R=d$ syndrome-extraction rounds using \texttt{Stim}~\cite{gidney2021stim}. Samples are generated online under the circuit-level depolarizing noise model, with the distance-specific physical error rates and training budgets listed in Table~\ref{tab:surface_training}. Following the logical-error-classification objective of SI~\ref{subsec:transformer_train}, the \ntu-Transformer is trained on binary detector-event histories and supervised by the final logical observable label through the loss $\ell(\hat{\bm{y}}_d,\bm{y}_d)$ in Eq.~(\ref{eq:distill}). To provide process supervision, we construct prefix syndromes by masking all detector events after each intermediate round and decode these truncated histories with a correlated PyMatching teacher, as described in SI~\ref{subsec:transformer_train}. The resulting intermediate binary cross-entropy loss is the second term of Eq.~(\ref{eq:distill}), weighted by the time-progressive coefficients $w_r$ and combined with the final-label loss. The distillation coefficient $\beta(s)$ of Eq.~(\ref{eq:distill}) is annealed from 8.0 to 0.2 over the early training stage, so that intermediate supervision dominates the cold-start phase but does not constrain the late-stage decoder to imitate the teacher.

We use stabilizer dropout, a QEC-specific variant of dropout~\cite{srivastava2014dropout} that randomly masks stabilizer detector events during training, as a regularizer for the surface-code \ntu-Transformer. During the first 50\% of training, the mask probability is fixed at $p_{\mathrm{mask}}=0.8$ with drop ratio $p_{\mathrm{drop}}=0.5$. Both quantities are then linearly decayed to zero between 50\% and 80\% of training and remain disabled for the final 20\%. Optimization uses AdamW with bfloat16 mixed precision, a 100-step linear warmup, cosine learning-rate decay to 5\% of the peak value, and global gradient clipping with maximum norm $1.0$. Weight decay $10^{-2}$ is applied only to structural projection weights, while biases, normalization parameters, embedding tables, logical query tokens, and readout heads are excluded from weight decay. Training is parallelized with PyTorch distributed data parallelism, and online teacher-label generation is performed by CPU workers.

For cross-distance transfer, all shape-compatible learnable parameters are loaded from the source checkpoint, including the embedding tables, recurrent layers, Transformer blocks, readout layers, and logical query parameters. Distance-dependent buffers are not transferred. These include coordinate maps, gather indices, validity masks, same-type and cross-type neighbour indices, attention masks, and other topology buffers. They are regenerated from the target-distance circuit before fine-tuning. The optimizer state is reset after transfer, and a new warmup--cosine schedule is started for the target run.

\begin{table}[t]
\centering
\small
\setlength{\tabcolsep}{4.5pt}
\renewcommand{\arraystretch}{1.08}
\caption{
Training configurations for the surface-code \ntu-Transformer experiments.
All runs use online \texttt{Stim} sampling with $R=d$ syndrome-extraction rounds.
}
\label{tab:surface_training}
\begin{tabular}{ccccccc}
\toprule
Distance $d$ & Rounds $R$ & Peak LR & $p_{\mathrm{train}}$ & $p_{\mathrm{eval}}$ & Local batch & Max steps \\
\midrule
7  & 7  & $2\times10^{-5}$ & 0.005 & 0.005 & 32 & 100,000 \\
11 & 11 & $2\times10^{-5}$ & 0.005 & 0.005 & 16 & 150,000 \\
15 & 15 & $2\times10^{-5}$ & 0.005 & 0.005 & 8  & 200,000 \\
19 & 19 & $2\times10^{-5}$ & 0.005 & 0.005 & 4  & 250,000 \\
23 & 23 & $2\times10^{-5}$ & 0.007 & 0.007 & 2  & 150,000 \\
25 & 25 & $2\times10^{-5}$ & 0.007 & 0.007 & 2  & 150,000 \\
\bottomrule
\end{tabular}
\end{table}

\smallskip
\noindent\textbf{Baseline methods.}
We benchmark the \ntu-Transformer against three non-trainable decoding baselines for the rotated surface code: standard PyMatching, Correlated PyMatching, and Tesseract. These baselines span the most relevant accuracy--complexity spectrum for circuit-level surface-code decoding: a fast graphlike MWPM decoder, a stronger correlation-aware matching decoder, and a search-based most-likely-error reference.

\smallskip
\noindent\emph{Standard PyMatching.}
We use PyMatching as the standard minimum-weight perfect matching baseline. For each circuit, we first extract a graph-like detector error model using \texttt{stim.Circuit.detector\_error\_model(decompose\_errors=True)}. The resulting DEM is loaded with \texttt{pymatching.Matching.from\_detector\_error\_model}, and decoding is performed in batch mode with \texttt{decode\_batch}. This baseline treats the decomposed error mechanisms as independent graph-like edges, following the standard MWPM assumption~\cite{edmonds1965paths,higgott2025sparse}. It therefore provides the canonical fast matching reference for surface-code decoding, but it does not explicitly use the residual correlations introduced when non-graph-like physical faults are decomposed into graph-like components.

\smallskip
\noindent\emph{Correlated PyMatching.}
We additionally evaluate the correlated matching mode implemented in PyMatching. The decoder is constructed from the same decomposed Stim DEM, but with \texttt{enable\_correlations=True} during both graph construction and batch decoding.
We include this decoder as a stronger matching-based baseline that partially accounts for circuit-level error correlations while remaining within the MWPM decoding paradigm.

\smallskip
\noindent\emph{Tesseract.}
Tesseract~\cite{beni2025tesseractdecoder} is included as an accuracy-oriented search-based reference. It is provided with the same detector error model and evaluated on the same held-out detector-event samples as the other baselines. Because its search procedure is substantially more computationally demanding than matching-based decoding at large distances, we use Tesseract primarily as a high-accuracy reference rather than as a scalable throughput baseline. In our experiments, direct Tesseract decoding was feasible for $d=3,\ 5,\ 7,\ 9$ at $p=0.003$; we use these results to extrapolate its performance to larger distances, including $d=19$.

\subsection{Experimental setup for BB code}\label{subsec:exp_bb}
\smallskip
\noindent\textbf{Experimental protocol.}
For the BB-code experiments, detector events are generated online with \texttt{Stim} using the BB memory circuits and the CNOT-layer syndrome-extraction schedule. A uniform circuit-level depolarizing error rate $p$ is applied to Clifford gates, reset operations, measurements, and pre-round data depolarization. The code-specific learning rates, batch sizes, gradient-accumulation factors, and training horizons are listed in Table~\ref{tab:bb_training}. We evaluate both \ntu-Transformer and \ntu-Neural-BP on the same detector-event distribution.

The \ntu-Transformer is trained as a logical error classifier using the final logical observable labels. Unlike the surface-code experiments, we do not use MWPM-based process distillation for BB codes, because the BB detector-error structure is non-graphlike and does not provide the same matching-based prefix teacher. We therefore set the distillation coefficient to zero. Stabilizer dropout is also disabled for Transformer runs, with $p_{\mathrm{mask}}=p_{\mathrm{drop}}=0$. Optimization uses AdamW with bfloat16 mixed precision, a 100-step warmup, cosine decay to 5\% of the peak learning rate, and global gradient clipping with maximum norm $1.0$. The same selective weight-decay strategy as in the surface-code Transformer is used: structural projection weights receive weight decay $10^{-2}$, while biases, normalization parameters, topological embeddings, logical query tokens, and readout heads are excluded.

For \ntu-Neural BP, the \texttt{Stim} detector-error model is converted into a sparse representation consisting of a detector-incidence map and a logical-observable map. The model is trained in the physical-error-inference paradigm, where the prediction variables are the error variables defined by this detector-error representation. The objective combines a focal loss with $\alpha=0.95$ and $\gamma=2.0$ with a syndrome-consistency loss weighted by $\lambda_{\mathrm{syn}}=0.2$. The BP unrolling depth is fixed to $N_{\mathrm{iter}}=2d$. Optimization uses AdamW in float32, a 100-step warmup, cosine decay to 5\% of the peak learning rate, global gradient clipping with maximum norm 1.0, and uniform weight decay $10^{-4}$. No edge-level or message-level dropout is applied.

Transfer from $[\![72,12,6]\!]$ to $[\![144,12,12]\!]$ follows the same separation between learnable parameters and code-dependent buffers. For \ntu-Transformer, all shape-compatible learnable parameters are loaded, while coordinate maps, masks, neighbor indices, and other topology buffers are regenerated for the target code. For \ntu-Neural-BP, the trainable message-passing parameters are reloaded, while the sparse graph tensors, priors, detector-incidence map, and logical-observable map are replaced by their target-code versions. In both cases, the optimizer state is reset before target-side fine-tuning.

\begin{table}[t]
\centering
\small
\setlength{\tabcolsep}{4.2pt}
\renewcommand{\arraystretch}{1.08}
\caption{
Training configurations for the BB-code experiments.
The \ntu-Transformer is trained for logical error classification, while \ntu-Neural-BP is trained for physical error inference on the detector-error representation.
}
\label{tab:bb_training}
\begin{tabular}{cccccccccc}
\toprule
Code & Decoder & Rounds ($R$) & BP iter. & Peak LR & $p_{\mathrm{train}}$ & $p_{\mathrm{eval}}$ & Local batch & Grad. accum. & Max steps \\
\midrule
$[\![72,12,6]\!]$   & \ntu-Transformer & 6  & -- & $5\times10^{-4}$ & 0.005 & 0.005 & 512 & 1 & 50,000 \\
$[\![144,12,12]\!]$ & \ntu-Transformer & 12 & -- & $5\times10^{-4}$ & 0.005 & 0.005 & 256 & 4 & 30,000 \\
$[\![72,12,6]\!]$   & \ntu-Neural-BP     & 6  & 12 & $3\times10^{-4}$ & 0.005 & 0.005 & 512 & 1 & 15,000 \\
$[\![144,12,12]\!]$ & \ntu-Neural-BP     & 12 & 24 & $3\times10^{-4}$ & 0.005 & 0.005 & 512 & 1 & 15,000 \\
\bottomrule
\end{tabular}
\end{table}

\smallskip
\noindent\textbf{Baseline methods.}
For BB-code experiments, we benchmark against BP+OSD~\cite{roffe2020decoding,panteleev2021degenerate} and RelayBP~\cite{muller2025improved}, whose decoding principles are reviewed in SI~\ref{subsec:pre_bb}. Both baselines are training-free and operate on the same detector-error representation used for the corresponding BB evaluations. Here we specify only their concrete implementations and hyperparameters.

\smallskip
\noindent\emph{BP+OSD.}
We use BP+OSD as a standard classical decoder for qLDPC codes~\cite{panteleev2021degenerate,roffe2020decoding}. The Stim DEM is converted into a sparse bipartite parity-check matrix $H$ together with the corresponding per-mechanism priors, and the decoder is instantiated using \texttt{ldpc.BpOsdDecoder} in sum-product mode with combination-sweep ordered-statistics decoding (\texttt{osd\_method="osd\_cs"}, \texttt{osd\_order=10}). The role of the OSD post-processing stage in restoring a syndrome-consistent estimate when raw BP fails is described in SI~\ref{subsec:pre_bb}.

\smallskip
\noindent\emph{RelayBP.}
We additionally evaluate RelayBP~\cite{muller2025improved}, a recent BP-based heuristic whose two-stage relay schedule for escaping degenerate trapping-set failures in quantum LDPC decoding is summarized in SI~\ref{subsec:pre_bb}. We use the released implementation accompanying Ref.~\cite{muller2025improved} with its default memory-factor distribution.
We report two RelayBP configurations. The original RelayBP setting uses 80 BP iterations in the first stage and 60 BP iterations in the second stage, with 300 parallel second-stage relay branches. The depth-aligned RelayBP-$2d$ setting uses $2d$ BP iterations in the first stage and $2d$ BP iterations in each second-stage relay branch, again with 300 parallel relay branches. The RelayBP-$2d$ configuration is aligned with the \ntu-Neural-BP unrolling depth, while the original RelayBP setting provides a higher-complexity reference following the released decoder configuration. In both cases, decoding terminates once a syndrome-consistent error configuration is found or all scheduled relay branches have been exhausted.

\section{Extended experimental results}\label{apx:results}
This section collects additional simulation results that support the main text. SI~\ref{subsec:ext_scaling} extends the training-compute scaling analysis to larger code distances, SI~\ref{subsec:coldstart} diagnoses the cold-start collapse of scratch-initialized training, SI~\ref{subsec:struct_ablation} ablates the two scale-invariant structural components of \ntu-Transformer, and SI~\ref{subsec:underconverged} shows that cross-distance transfer remains effective from under-converged source checkpoints.

\subsection{Extended training compute scaling}\label{subsec:ext_scaling}
\begin{figure}
    \centering
    \includegraphics[width=0.5\linewidth]{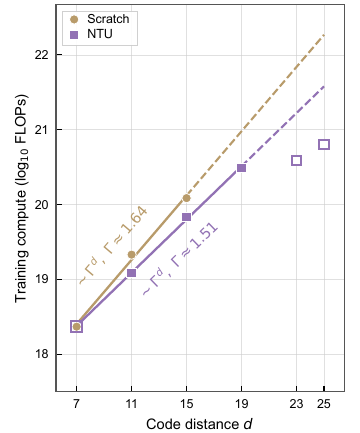}
    \caption{\textbf{Extended training-compute scaling of \ntu-Transformer on rotated surface codes.}
Training compute is measured as the total FLOPs required for the decoder to exceed the PyMatching reference accuracy. Filled markers denote runs performed under the controlled protocol used in the main text, with $p_{\mathrm{train}}=5\times10^{-3}$. Dashed lines show exponential scaling fits for scratch training and \ntu, with fitted bases $\Gamma=1.64$ and $\Gamma=1.51$, respectively. Open squares indicate large-distance stress-test runs at $d=23$ and $d=25$, performed with $p_{\mathrm{train}}=7\times10^{-3}$ and a larger hardware configuration. These two points are shown to indicate feasibility beyond the controlled range, but are not included in the controlled scaling fit.}
    \label{fig:extended-compute-scaling}
\end{figure}
In the main text, the training-compute scaling analysis was performed under a controlled protocol for distances up to $d=19$. In that setting, all runs used the same training physical error rate $p_{\mathrm{train}}=5\times 10^{-3}$, the same evaluation criterion, and the same hardware class described in the remark of SI~\ref{apx:exp_details}. This controlled setting was used to estimate the scaling trend reported in Fig.~\ref{fig:scheme}c.

To further probe the large-distance regime, we extended the transfer ladder to $d=23$ and $d=25$. These two runs required a modified experimental setting because the original hardware configuration became memory-limited at these distances. Accordingly, the large-distance stress tests were performed on 16 NVIDIA H200 GPUs with 140 GB memory per GPU, and the training physical error rate was increased to $p_{\mathrm{train}}=7\times 10^{-3}$. This choice increases the density of nontrivial detector-event patterns and empirically reduces the number of training samples required to exceed the PyMatching reference. However, $p_{\mathrm{train}}=7\times 10^{-3}$ is close to the high-noise transition region of this training setup, and models trained at this noise level do not necessarily give the best inference performance at substantially lower physical error rates. We therefore treat the $d=23$ and $d=25$ runs as out-of-protocol scaling probes rather than as points generated under the same controlled conditions as $d\le19$.

In SI Fig.~\ref{fig:extended-compute-scaling}, these two large-distance points are shown with open square markers to distinguish them from the controlled $p_{\mathrm{train}}=5\times10^{-3}$ runs. The measured training compute values are $3.08\times10^{20}$, $3.88\times10^{20}$, and $6.30\times10^{20}$ FLOPs for $d=19$, $d=23$, and $d=25$, respectively. The corresponding wall-clock training costs are 896, 768, and 1048 GPU-hours, respectively. Because the $d=23$ and $d=25$ experiments use both a different noise distribution and a different hardware platform, their wall-clock times are not directly comparable to the $d\le19$ runs. In particular, the larger GPU memory and increased parallelism can reduce wall-clock cost even when the profiled FLOP count increases. We thus use the $d=23$ and $d=25$ measurements as reference points for large-distance feasibility and not as part of the controlled scaling fit.

\subsection{Cold-start in scratch training}\label{subsec:coldstart}
\begin{figure}
    \centering
    \includegraphics[width=0.9\linewidth]{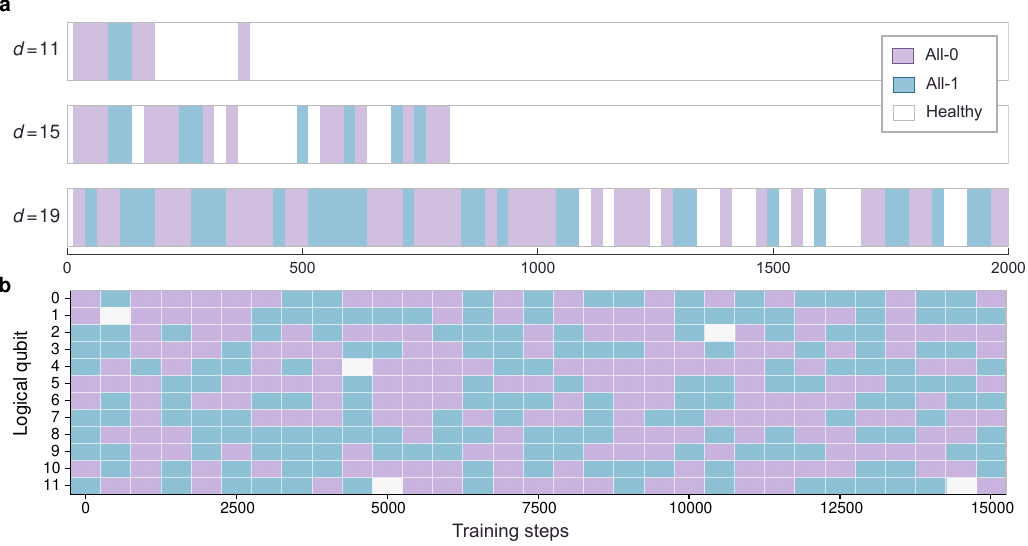}
    \caption{\textbf{Cold-start mode-collapse diagnostics for scratch-initialized \ntu-Transformer training.}
\textbf{a,} Surface-code logical classification. Each horizontal bar tracks the output state of a scratch-initialized decoder during early training at distances $d=11,15,19$. Checkpoints are classified as all-0 collapse, all-1 collapse, or healthy according to whether the thresholded predictions on the validation batch are constant or non-constant. As the distance increases, the decoder spends longer intervals in constant-output states, indicating that the 50\% accuracy plateau is dominated by trivial prediction modes rather than unbiased random guessing.
\textbf{b,} BB-code logical classification. The same diagnostic is applied separately to the 12 logical outputs of the BB-code decoder. Each row corresponds to one logical qubit, and each column corresponds to a training checkpoint. The frequent alternation between all-0 and all-1 states shows that cold-start collapse also occurs at the level of individual logical heads in the multi-logical-qubit setting.}
    \label{fig:cold_start}
\end{figure}
The cold-start plateau in Fig.~\ref{fig:surface}c is not fully explained by accuracy alone. For the single-logical-qubit surface-code memory task at $p=0.005$, the logical labels are approximately balanced between 0 and 1. A decoder that makes random binary predictions and a decoder that collapses to a constant output can both remain close to 50\% block accuracy. This makes it difficult to distinguish slow feature learning from a trivial output mode using accuracy alone.

A common way to make early training easier is to begin from a smaller physical error rate, where nontrivial logical labels are rarer and the positive-label fraction is reduced~\cite{senior2025scalable,gu2026scalable}. This forms a curriculum over the label distribution, but it does not remove the intrinsic optimization difficulty of training a high-capacity decoder with tens of millions of parameters at large distance. Here, instead of changing the training distribution, we directly diagnose the scratch-trained outputs at the same $p=0.005$ setting. At each checkpoint, we threshold the predicted logical logits on a held-out validation batch and classify the checkpoint as all-0 collapse, all-1 collapse, or healthy, depending on whether all validation samples are assigned to a constant logical value.

SI Fig.~\ref{fig:cold_start} shows that the scratch-training plateau is dominated by constant-output modes rather than unbiased random guessing. As the distance increases, the scratch-initialized \ntu-Transformer spends longer intervals switching between all-0 and all-1 predictions instead of learning a meaningful syndrome-to-logical mapping. The same behaviour is also observed in BB-code logical classification, where different logical heads can independently become trapped in constant-output states. This failure mode is more pronounced for BB codes because the readout must assign multiple logical queries to different distributed detector features while optimizing several logical-output losses simultaneously. \ntu~initialization reduces this failure mode by providing a pre-aligned local perception backbone, so target fine-tuning mainly recalibrates the enlarged global decision layer instead of first escaping a trivial constant predictor.

\subsection{Structural ablation of \ntu-Transformer}\label{subsec:struct_ablation}
\begin{figure}
    \centering
    \includegraphics[width=0.95\linewidth]{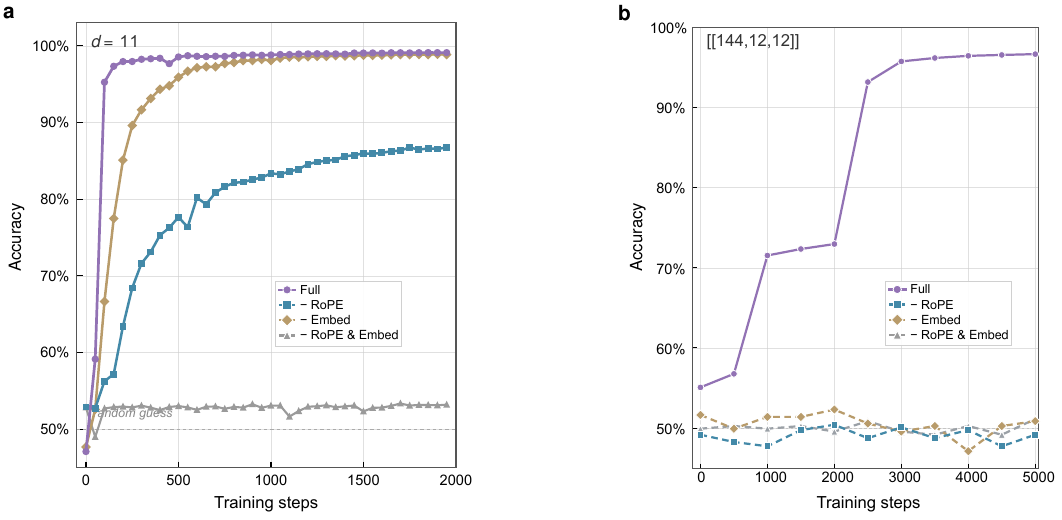}
    \caption{\textbf{Structural ablation of \ntu-Transformer at $p=0.005$.}
\textbf{a,} Surface-code transfer to the $d=11$ target task. Removing either geometry-aware RoPE or the topology-driven embedding module slows early fine-tuning, while removing both components leaves the transferred model close to the random-guess baseline within the plotted training window. Here, ``Embed'' denotes the topology-driven input embedding used to align detector features across code sizes.
\textbf{b,} BB-code transfer to the $[\![144,12,12]\!]$ target task. In this non-local geometry, transfer is more sensitive to structural alignment: only the full model shows a sustained escape from the cold-start regime, whereas the RoPE-ablated, embedding-ablated, and double-ablated variants remain near random guessing. The accuracy here is the averaged over all 12 logical qubits.
}
    \label{fig:ablation}
\end{figure}
\smallskip
\noindent\textbf{Surface-code ablation.}
We first perform the architecture ablation on the rotated surface code, where the local planar geometry makes the two structural transfer channels especially transparent, as shown in SI Fig.~\ref{fig:ablation}. The transfer task initializes a $d=11$ target decoder from a pretrained $d=7$ \ntu-Transformer under circuit-level depolarizing noise with $p_{\rm train}=p_{\rm eval}=5\times 10^{-3}$. We focus on the first $2{,}000$ target fine-tuning steps, so the comparison should be interpreted as a transfer-efficiency test rather than a comparison of ultimate trainability given unlimited optimization.

As shown in SI Fig.~\ref{fig:ablation}a, the Full model exhibits the fastest lift-off from the random-guess baseline, showing that physical-coordinate RoPE and the local one-ring stem together provide a strongly aligned initialization for the larger lattice. Ablating either component slows this early fine-tuning process. Removing RoPE weakens the relative geometric phase information in the spatial attention maps, while removing the local stem weakens the consistency between the discrete input embedding vocabulary and the same physical surface-code neighborhood after scaling. Nevertheless, either structural channel alone is sufficient to retain a nontrivial transfer trajectory on the planar surface code. Only the double ablation remains close to random guessing over the fixed early horizon, indicating that when both physical alignment channels are removed, the transferred checkpoint no longer provides a measurable early acceleration over the cold-start regime.

Thus, for the surface code, RoPE and the local stem should be understood primarily as transfer accelerators rather than strict determinants of final model capacity. Their role is to preserve scale-invariant local physical priors so that target-side fine-tuning starts from a nontrivial basin instead of spending substantial optimization compute escaping the cold-start plateau.

\smallskip
\noindent\textbf{BB-code ablation.}
We next perform the same type of architecture ablation on BB-code transfer, where the quasi-cyclic non-local topology makes the consistency of the structural alignment more stringent. To isolate the contribution of the two scale-invariant structural components of \ntu-Transformer, namely the geometry-aware rotary positional encoding and the polynomial-induced topological embedding, we conduct a $2\times2$ ablation on the $[\![72,12,6]\!]\to[\![144,12,12]\!]$ transfer task, as shown in SI Fig.~\ref{fig:ablation}b. All four runs share the optimization protocol of SI~\ref{subsec:exp_bb} and are trained for $5{,}000$ fine-tuning steps, which is sufficient to discriminate sustained lift-off from the cold-start basin.

The two ablated implementations are defined as follows. The \emph{code-size-scaled torus RoPE} parameterizes the inverse frequencies as $\mathrm{inv\_freq}_i \propto 2\pi/\ell$ along the first torus axis and analogously $\mathrm{inv\_freq}_j \propto 2\pi/m$ along the second. Placing the code-size constants $\ell$ and $m$ in the denominator causes the same relative offset $\Delta i$ to produce different RoPE phases on $[\![72,12,6]\!]$ ($\ell=6$) and $[\![144,12,12]\!]$ ($\ell=12$). The Full model instead uses the Cartesian, code-size-independent RoPE introduced in SI~\ref{apx:transformer}, $\mathrm{inv\_freq}_k=\Theta^{-k/\mathrm{quarter\_dim}}$ with $\Theta=100$, whose inverse-frequency buffer is identical across both codes.

The \emph{torus $L_2$-nearest embedding} replaces the polynomial-induced offset set of SI~\ref{apx:transformer} (which yields $K_{\rm same}=12$ same-type and $K_{\rm cross}=9$ cross-type neighbors derived from the off-diagonal terms of $AA^{T}$, $BB^{T}$, and $AB$) with a geometric heuristic that selects each stabilizer's $K=6$ torus-$L_2$-nearest same-type neighbors. The resulting neighbor set includes pairs that share no common data qubit on the BB Tanner graph while omitting genuine polynomial-induced neighbors at larger $L_2$ distance, so the same embedding index no longer encodes the same physical relation across the two code sizes.

In contrast to the surface-code case, breaking either alignment is sufficient to remove the measurable early transfer advantage on BB codes. The RoPE ablated, embedding ablated, and double-ablated configurations all remain close to the $50\%$ random-guess plateau within the $5{,}000$-step fine-tuning horizon, while only the Full model exhibits sustained lift-off. This does not imply that the target architectures are untrainable given enough optimization compute; rather, it shows that the specific cross-code transfer acceleration provided by \ntu~depends on preserving both the positional phase alignment and the polynomial neighborhood alignment in this non-local quasi-cyclic topology.

\subsection{Transfer from under-converged source checkpoints}\label{subsec:underconverged}
\begin{figure}[t]
\centering
\includegraphics{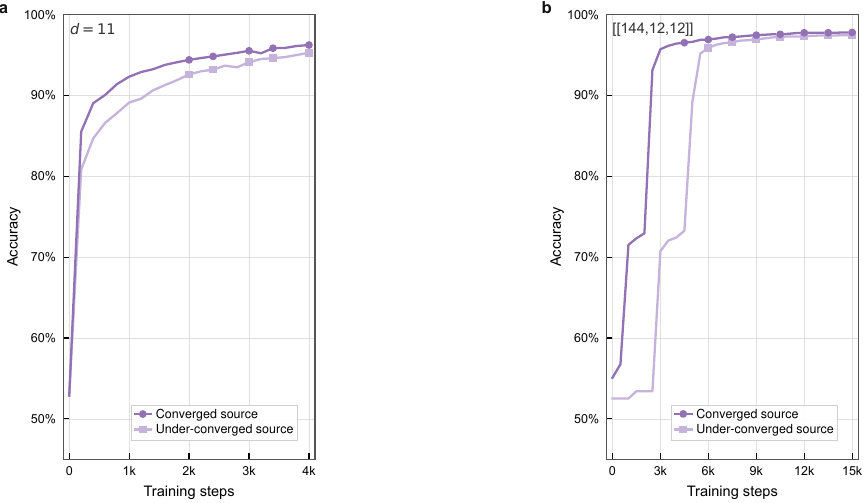}
\caption{
\textbf{Transfer from under-converged source checkpoints.}
\textbf{a,} Rotated surface-code transfer from a $d=7$ source checkpoint to the $d=11$ target task.
The source-task accuracies of the converged and under-converged checkpoints are 98\% and 97\%, respectively.
Although the under-converged source gives a lower initial target accuracy, its fine-tuning trajectory rises on a comparable time scale and progressively approaches the converged-source transfer.
\textbf{b,} BB-code transfer from $[\![72,12,6]\!]$ source checkpoints to the $[\![144,12,12]\!]$ target task.
The source-task accuracies of the converged and under-converged checkpoints are 88\% and 78\%, respectively.
Despite the lower source convergence, both target trajectories escape the 50\% random-guess baseline within the early fine-tuning stage and approach the same accuracy range after continued training.
Together, these experiments show that effective \ntu\ framework does not require complete source-task convergence: an under-converged source model can already provide a useful scale-invariant initialization for larger target codes, with the lower initial target accuracy largely compensated by subsequent target-side fine-tuning.
}
\label{fig:transfer_uncompleted}
\end{figure}

A practical question for the \ntu\ framework is whether the source model must be fully converged before it can provide a useful initialization for a larger target code. If transferable structural priors emerge before source-task saturation, the pretraining compute required to initiate cross-distance transfer can be reduced. We test this directly on both representative code families by transferring from source checkpoints taken at different convergence levels along the same pretraining trajectory, while keeping the target architecture, optimizer schedule, noise setting, and fine-tuning protocol fixed.

As shown in SI Fig.~\ref{fig:transfer_uncompleted}, under-converged source checkpoints remain effective for both the rotated surface-code and BB-code transfers. In the surface-code experiment, a lower-accuracy $d=7$ source checkpoint transfers successfully to the $d=11$ target task. In the BB-code experiment, an under-converged $[\![72,12,6]\!]$ source checkpoint similarly transfers to the larger $[\![144,12,12]\!]$ target task. In both cases, the less-converged source starts from a lower target accuracy than the more-converged source, indicating that the transferred representation is not yet fully calibrated. However, this initial deficit does not prevent target-side learning: the fine-tuning curves rise on comparable time scales, avoid the random-guess plateau, and approach similar accuracy ranges after continued training.

These results indicate that the \ntu\ framework does not rely on complete source-task saturation. The transferable ingredients appear to be local, scale-invariant topological priors encoded in the perception backbone, which emerge before the source task reaches its final accuracy. Full source convergence mainly improves the initial target calibration, whereas the remaining mismatch introduced by an under-converged source checkpoint can be absorbed by continued target-side fine-tuning. This further reduces the effective pretraining burden of the \ntu\ framework, since cross-distance transfer can be initiated once a nontrivial local perception backbone has formed rather than only after exhaustive source-task optimization.

\end{document}